\documentclass[a4paper,11pt]{article}

\usepackage{amsmath,amssymb}
\usepackage[page]{appendix}
\usepackage{authblk}
\usepackage{bbm}
\usepackage{bm}
\usepackage{cite}
\usepackage{comment}
\usepackage{enumitem}
\usepackage{geometry}
\usepackage{graphicx}
\usepackage[scr=boondox]{mathalpha}
\usepackage{mathtools}
\usepackage{physics}
\usepackage{slashed}
\usepackage{subcaption}
\usepackage{subfiles}
\usepackage{upgreek}
\usepackage{empheq} 
\usepackage{xcolor}
\usepackage{forest}
\usepackage{cancel}
\usepackage{fontawesome5}

\usepackage[most]{tcolorbox} 

\usepackage{listings}
\lstdefinestyle{datafile}{
  basicstyle={\ttfamily\footnotesize},
  numbers=left,
  numberstyle=\tiny\color{gray},
  numbersep=5pt,
  breaklines=true,
  frame={lines},
  rulecolor=\color{black},
  framerule=0.5pt,
  columns=flexible,
  tabsize=2,
  keywordstyle={},
  commentstyle=\color{gray},
  morecomment=[l]{\#}
}

\usepackage[colorlinks=true,breaklinks=true]{hyperref}
\hypersetup{
allcolors=[rgb]{0.0 0.0 0.6},
linkcolor=[rgb]{0.0 0.0 1}
}

\colorlet{lightblue}{blue!30}
\colorlet{lightred}{red!30}

\newcommand{\Born}{{\rm Born}}
\newcommand{\elas}{{\rm elas}}
\newcommand{\fin}{{\rm fin}}
\newcommand{\im}{\mathbbm{i}}
\newcommand{\irr}{{\rm irr}}
\newcommand{\med}{{\rm med}}
\newcommand{\vrel}{v_{\rm rel}}
\newcommand{\rgl}{{\rm rgl}}
\newcommand{\cl}{{\rm cl}}
\newcommand{\xcl}{x_{\rm cl}}
\newcommand{\bxcl}{\bar{x}_{\rm cl}}
\newcommand{\sym}{\mathscr{C}}

\newcommand{\gammaE}{{\gamma_{\rm E}}}

\newcommand{\T}{{\scriptscriptstyle{\rm T}}}
\newcommand{\U}{{\scriptscriptstyle{\rm U}}}

\newcommand{\V}{{\scriptscriptstyle{\rm V}}}
\newcommand{\WKB}{{\scriptscriptstyle{\rm WKB}}}

\newcommand{\RR}{\mathbbm{R}}

\AtBeginEnvironment{appendices}{\crefalias{section}{appendix}}
\numberwithin{equation}{section}

\usepackage[capitalise]{cleveref}
\crefformat{pluralequation}{#2{\color{black} Eqs.~(}#1{\color{black} )}#3}
\Crefformat{pluralequation}
{#2{\color{black} Equations~(}#1{\color{black} )}#3}

\title{\bf 
Variable-phase methods and data\\
for long-range scattering}
\author[1,2]{\Large
Marcos M. Flores\footnote{marcos.flores@fys.uio.no}}  
\author[1]{Kalliopi Petraki\footnote{kalliopi.petraki@phys.ens.psl.eu}}
\author[1]{Thomas Pignard\footnote{thomas.pignard@phys.ens.psl.eu}}
\affil[1]{\large\it
Laboratoire de Physique de l'École Normale Supérieure, 
ENS, 
Université PSL, 
CNRS, 
Sorbonne Université, 
Université Paris Cité,
Paris, F-75005, France}
\affil[2]{\large\it 
Department of Physics, University of Oslo, Box 1048, N-0316 Oslo, Norway }

\begin{document}
\maketitle
\begin{abstract}
Long-range interactions play an important role in dark-matter phenomenology, affecting the relic abundance, indirect-detection signals, and structure formation on galactic scales. Reliable predictions often require nonperturbative solutions of the Schr\"odinger equation over broad parameter ranges and many partial waves. We present a numerical framework based on the variable-phase method, which recasts the radial Schr\"odinger equation as a system of first-order equations and allows both regular and irregular solutions to be computed efficiently. The former determine phase shifts and Sommerfeld factors, while the latter provide additional information relevant for the self-consistent unitarization of contact-like inelastic interactions. We discuss the numerical implementation, integration-domain selection, partial-wave convergence, and comparisons with analytic limits. We release the code together with tables for the Yukawa potential containing phase shifts, Sommerfeld factors, elastic and angularly weighted cross sections, and the irregular-solution information relevant for unitarization.  These resources facilitate dark-matter studies without requiring repeated computationally intensive scans.
\end{abstract}

\clearpage
\noindent\rule{\textwidth}{0.4pt}
\tableofcontents
\noindent\rule{\textwidth}{0.4pt}

\clearpage
\section{Introduction}\label{sec:Intro}

An interaction behaves effectively as long-ranged when its range is comparable to or larger than the de Broglie wavelength of the interacting particles. Accurate calculations of scattering in this regime are important across many areas of physics, from atomic and nuclear systems to dark-matter (DM) phenomenology. In multi-TeV DM models, long-range interactions can substantially modify the thermal relic abundance through the Sommerfeld effect and radiative bound-state formation~\cite{Hisano:2006nn,vonHarling:2014kha}, as well as annihilation rates relevant for indirect searches~\cite{Hisano:2003ec,Hisano:2004ds,Pospelov:2008jd,March-Russell:2008klu}. The importance of these effects has been demonstrated in model-dependent studies, and is supported by model-independent unitarity arguments~\cite{Baldes:2017gzw,Flores:2024sfy}. In self-interacting dark-matter (SIDM) scenarios, light mediators can generate strongly velocity-dependent elastic cross sections that affect structure formation on galactic scales~\cite{Spergel:1999mh,Loeb:2010gj,Tulin:2017ara}. When these long-range effects become nonperturbative, scattering must be treated by solving the Schr\"odinger equation, which resums repeated interactions between the two particles to all orders (see e.g.~\cite{Petraki:2015hla}).

For the Coulomb potential, the Schrödinger equation admits analytic solutions, but most other physically relevant potentials, including the ubiquitous Yukawa potential, must be treated numerically. The standard approach is to solve the second-order Schrödinger equation subject to the appropriate boundary conditions at small and large distances. The resulting wave functions determine the phase shifts relevant for elastic scattering~\cite{Tulin:2013teo}, the Sommerfeld factors multiplying short-distance inelastic processes such as annihilation~\cite{Cassel:2009wt}, and overlap integrals entering radiative processes such as bound-state formation~\cite{Petraki:2015hla,Petraki:2016cnz}. Self-consistent unitarization of contact-like inelastic interactions may require, beyond the phase shift and Sommerfeld factor, additional information from the same Schr\"odinger problem that can be extracted from the irregular solution~\cite{Blum:2016nrz,Flores:2025qit,Parikh:2024mwa}. Such calculations become computationally demanding when broad parameter regions must be scanned. The difficulty is greater near resonances associated with bound states close to threshold, and in regimes where many partial waves contribute, as occurs in SIDM~\cite{Tulin:2013teo} and in some bound-state-formation processes~\cite{Oncala:2019yvj,Flores:2026yay,Binder:2023ckj,Beneke:2024nxh}.

Several related computational methods and public tools address complementary aspects of these problems. Model-specific WIMP calculations include Sommerfeld-enhanced annihilation through coupled-channel Schr\"odinger equations, with implementations such as DarkSE within the DarkSUSY ecosystem~\cite{Hryczuk:2010zi,Hryczuk:2011tq,Beneke:2016ync,Bringmann:2018lay}. Other public tools include DRAKE, which employs Hulth\'en-based approximations for Sommerfeld-enhanced annihilation~\cite{Binder:2021bmg}, and DM@NLO and SE+BSF4DM, which incorporate Sommerfeld and bound-state effects in selected MSSM or colored dark-sector settings, exploiting analytic formulae available for Coulombic QED or QCD potentials~\cite{ElHedri:2016onc,Harz:2023llw,Becker:2025vgq}. For SIDM, numerical calculations and practical parametrizations of Yukawa scattering have been developed for halo phenomenology~\cite{Loeb:2010gj,Tulin:2013teo}, with the semiclassical regime implemented in CLASSICS~\cite{Colquhoun:2020adl}; more recently, \texttt{pykawa} provides numerical Yukawa phase shifts for self-interaction studies in astrophysical systems~\cite{Gilman:2026pyk}.

In this context, to our knowledge, no public resource for Yukawa-type potentials simultaneously provides phase shifts, Sommerfeld factors, and the additional information from the irregular solution needed for the self-consistent unitarization of contact-like inelastic interactions over a broad parameter space spanning the Born and resonant regimes and the transition to the semiclassical and Coulomb-like scattering. Both broad parameter coverage and the simultaneous availability of these quantities are important for exploring dark-sector phenomenology comprehensively, combining relic-density, indirect-detection, and structure-formation probes. This is particularly relevant given recent work showing that, when elastic or inelastic cross sections approach their partial-wave unitarity limits, they feed back into one another~\cite{Flores:2024sfy,Flores:2025qit}. A consistent unitarized calculation must treat the relevant elastic and inelastic dynamics together. The quantities released here provide the long-range inputs for such calculations, as illustrated by the companion studies of annihilation and bound-state-formation feedback in SIDM~\cite{Petraki:2026krj,Flores:2026jsf}.

In this work, we present a numerical strategy for solving the radial Schr\"odinger equation for real, local, central potentials, based on the variable-phase method~\cite{Calogero:1967var}. This formulation replaces the second-order radial equation by a system of first-order differential equations and is well suited to scans over broad regions of parameter space. The regular solution yields the phase shifts and Sommerfeld factors, while the same system can be integrated inward from large distances to construct the irregular solution and extract the information needed for unitarization. We describe how the integration domain is chosen across parameter space, derive analytic estimates of the partial-wave truncation error relevant for SIDM, and validate the numerical results against the Born, WKB, and Coulomb limits.

In addition, we release code that can be adapted to different potentials, together with tabulated data for the Yukawa potential, so that the results can be used directly in phenomenological studies. We provide tabulated phase shifts and Sommerfeld factors for $\ell\in[0,100]$, together with the total elastic, transfer, and viscosity cross sections obtained from them and the number of partial waves required for their convergence. For the $s$-wave, we additionally provide the information extracted from the irregular solution that is needed for the self-consistent unitarization of contact-like inelastic interactions. The phase shifts determine the elastic, transfer, and viscosity cross sections relevant for DM halo evolution, while the Sommerfeld factors encode the long-range modification of short-distance annihilation rates relevant for freeze-out and indirect detection. Although the full wave functions required for overlap integrals in bound-state formation and bremsstrahlung are not tabulated, they can be generated using the released code. 

This paper is organized as follows. \Cref{sec:Fundamentals} reviews the scattering formalism and defines the phase shifts, Sommerfeld factors, and the additional quantity relevant for unitarization. \Cref{sec:VPA} introduces the variable-phase approach and discusses the numerical computation of the regular and irregular solutions. \Cref{sec:Yukawa} applies the method to the Yukawa potential and compares the numerical results with analytic limits. \Cref{sec:ScatteringCS} presents the elastic, transfer, and viscosity cross sections obtained from the phase shifts and discusses the convergence of the corresponding partial-wave sums. We conclude in \cref{sec:Conclusions}. The code is publicly available at \href{https://github.com/mmfloresnm/YukawaVPS}{\faGithub\,\texttt{YukawaVPS}}, and the tabulated data are archived on Zenodo~\cite{Flores:2026data}. The repository contains the \texttt{Mathematica} notebooks used to generate the data, examples for importing and visualizing them, and a user guide.

\section{Fundamentals of scattering theory}
\label{sec:Fundamentals}

\subsection{Schrödinger equation}
\label{sec:Fundamentals_SchroedingerEq}

Our starting point is the time-independent Schrödinger equation for two nonrelativistic particles interacting through a real, local, and central potential $V(r)$,
\begin{equation}
\label{eq:SchroedingerEquation}
-\dfrac{\nabla^2}{2\mu}\,
\psi_{\vb{k}}(\vb{r})
+
V(r)
\psi_{\vb{k}}(\vb{r})
=
{\cal E}_{\vb{k}} 
\, \psi_{\vb{k}}(\vb{r}),
\end{equation}
where ${\cal E}_{\vb{k}} = \vb{k}^2/(2\mu)$, $\mu$ is the reduced mass, $\vb{k}=\mu \vb{v}_{\rm rel}$ is the momentum of one particle in the center-of-momentum (CM) frame and $\vb{v}_{\rm rel}$ is the relative velocity of the interacting particles. For a potential satisfying~\cref{eq:Potential_ShortRangeAssumption}, the scattering solution has the asymptotic behavior
\begin{equation}
\label{eq:WaveFunction_AsymptoteInfinity}
\psi_{\vb{k}}(\vb{r})
\ \underset{r\to\infty}{\simeq} \
\psi_{\vb{k}}^{\rm in} (\vb{r})
+
f_{\vb{k}}(\vartheta)\,
\frac{e^{\im kr}}{r},
\qquad
\cos\vartheta
\equiv
\hat{\vb{k}}\cdot\hat{\vb{r}}.
\end{equation}
The incoming wave is
\begin{align}
\label{eq:WaveFunction_IncomingWave}
\psi_{\vb{k}}^{\rm in} (\vb{r}) = 
\begin{cases}
e^{\im \vb{k}\cdot \vb{r}},
&\text{distinguishable particles},
\\
\dfrac{1}{\sqrt{2}}
\qty(
e^{\im \vb{k}\cdot \vb{r}} \pm 
e^{-\im \vb{k}\cdot \vb{r}}),
&\text{identical particles},    
\end{cases}
\end{align}
where, for identical particles, the sign is chosen such that the complete two-particle state has the exchange symmetry required by the particle statistics; it therefore depends on the exchange symmetry of the spin state and any other internal degrees of freedom.

\subsubsection*{Partial-wave analysis}

The wave function may be decomposed into partial waves as
\begin{align}
\label{eq:psi_PWanalysis}
\psi_{\vb{k}}(\vb{r}) 
&= \sum_{\ell=0}^{\infty}
(2\ell + 1)
P_\ell(\cos\vartheta)
\psi_{k, \ell}(r) ,
\end{align}
where $P_\ell$ is the Legendre polynomial of degree $\ell$. We define the reduced radial wave function, $u_{k,\ell}(r)$, by
\begin{align}
\label{eq:WavefunctionsRadial_psi-u}
u_{k,\ell} (r) \equiv k r\, \psi_{k,\ell} (r) ,
\end{align}
in terms of which the Schr\"odinger equation reduces, for each partial wave, to
\begin{align}
\label{eq:SchroedingerEq_Radial}
\left[
-\dfrac{1}{2\mu}\dv[2]{r} 
+\dfrac{\ell(\ell+1)}{2\mu \, r^2} 
+
V(r)
\right] u_{k,\ell} (r) 
= {\cal E}_{\vb{k}} \, u_{k,\ell} (r) .
\end{align}
We focus on potentials satisfying
\begin{equation}
\label{eq:Potential_ShortRangeAssumption}
\int_0^\infty \dd r\,r\,|V(r)|<\infty.
\end{equation}
This condition ensures that the potential is less singular than $1/r^2$ at the origin and decreases sufficiently rapidly at large distances.
Under this assumption, \cref{eq:SchroedingerEq_Radial} admits two linearly independent solutions with leading short-distance behavior,
\begin{align}
\label{eqs:RadialWaveFunctions_AsymptoteOrigin}
{\cal F}_{k,\ell}(r)
\ \underset{r\to 0}{\propto} \
r^{\ell+1} 
\quad \text{and} \quad
{\cal G}_{k,\ell}(r)
\ \underset{r\to 0}{\propto} \
r^{-\ell} .
\end{align}
${\cal F}_{k,\ell}$ and ${\cal G}_{k,\ell}$ belong to the regular and irregular families of solutions, respectively. A general radial solution can be expressed as a linear combination of the two linearly independent solutions,
\begin{equation}
u_{k,\ell}(r)
=
\alpha_{k,\ell}\,
{\cal F}_{k,\ell} (r)
+
\beta_{k,\ell} \,
{\cal G}_{k,\ell}(r).
\end{equation}
To treat distinguishable and identical particles with a common notation, we introduce the partial-wave symmetry factors
\begin{align}
\label{eq:SymmetryFactor}
\sym_\ell
\equiv
\begin{cases}
1,
& \text{distinguishable particles},
\\
2,
& \text{identical particles in surviving partial waves},
\end{cases}
\end{align}
and do not consider modes that are forbidden by exchange symmetry.

\subsection{Physical solution}

Solutions describing physical states must be regular at the origin. Since the full wave function is given by $\psi_{k,\ell}(r)=u_{k,\ell}(r)/(kr)$, regularity at $r=0$ excludes the irregular solution, setting $\beta_{k,\ell}=0$.

\subsubsection*{Phase shift}

The scattering amplitude admits a partial-wave decomposition analogous to that of the wave function,
\begin{equation} 
\label{eq:fk_PWanalysis}
f_{\vb{k}}(\vartheta)
=
\sum_{\ell=0}^{\infty}
(2\ell+1)\,
f_{k,\ell}\,
P_\ell(\cos\vartheta).
\end{equation}
Substituting \cref{eq:fk_PWanalysis} into the asymptotic scattering solution \eqref{eq:WaveFunction_AsymptoteInfinity}, the ratio of the outgoing and incoming coefficients in each partial wave is found to be $1+ 2\im k f_{k,\ell} / \sqrt{\sym_\ell}$. For a real potential, scattering is elastic and conservation of probability requires this quantity to have unit modulus,
\begin{equation}
\label{eq:PartialWaveAmplitude_Unitarity}
1+
\dfrac{2\im k}{\sqrt{\sym_\ell}}\,
f_{k,\ell}
=
e^{2\im\theta_\ell(k)},
\end{equation}
where the \emph{phase shift} $\theta_\ell(k) \in \RR$ is defined modulo $\pi$. It follows that
\begin{align}
\label{eq:fk_PW}
f_{k,\ell} 
=
\sqrt{\sym_\ell}
\, \dfrac{e^{2\im\theta_\ell(k)}-1}{2\im k}
= 
\sqrt{\sym_\ell}
\, \dfrac{1}{k}
\, e^{\im\theta_\ell(k)} \sin\theta_\ell (k) .
\end{align}
The corresponding physical radial solution therefore has the large-distance behavior
\begin{align}
\label{eq:PhysicalSolution_AsymptoteInfinity}
{\cal F}_{k,\ell}(r)
\ \underset{r\to \infty}{\sim} \
\sqrt{\sym_\ell}
\, \im^\ell
\, e^{\im \theta_\ell (k)}
\, \sin[kr - \ell\pi/2 + \theta_\ell(k)] .
\end{align}
The elastic scattering amplitude $f_{\vb{k}}(\vartheta)$ determines the differential cross section through\footnote{%
In our conventions, the wave functions for identical particles are (anti-)symmetrized as appropriate. \Cref{eq:Differential_CS_Def} then applies for both distinguishable and identical particles, over the full solid angle.}
\begin{align}
\label{eq:Differential_CS_Def}
\dv{\sigma}{\Omega} = 
|f_{\vb{k}}(\vartheta)|^2 .
\end{align}
Using \cref{eq:Differential_CS_Def,eq:fk_PWanalysis,eq:fk_PW} together with the orthogonality relations of the Legendre polynomials, \cref{eq:Legendre_Orthogonality}, the elastic cross section becomes
\begin{align}
\sigma^{\elas}(k) = \sum_{\ell = 0}^\infty \sigma_\ell^{\elas}(k),
\qquad
\sigma_\ell^{\elas}(k) = \sigma_\ell^{\U}(k)\, \sin^2\theta_\ell(k),
\end{align}
where we have defined the \emph{partial-wave unitarity cross section}~\cite{Flores:2024sfy,Flores:2025qit}
\begin{align}
\label{eq:sigma_Unitarity}
\sigma_\ell^{\U}(k) \equiv 
\sym_\ell \frac{4\pi}{k^2}(2\ell + 1) .
\end{align}

\subsubsection*{Sommerfeld factor} 
In the absence of a potential $V(r)$, the physical solution to \cref{eq:SchroedingerEq_Radial} normalized according to \cref{eq:PhysicalSolution_AsymptoteInfinity} is
\begin{align}
\label{eq:RadialWaveFunction_Free}
{\cal F}_{k,\ell}^{\rm free}(r) = 
\sqrt{\sym_\ell}
\, \im^\ell 
\, kr 
\, j_\ell(kr) ,
\end{align}
where $j_\ell$ is the spherical Bessel function of the first kind. At sufficiently small $r$,
\begin{align}
\label{eq:RadialWaveFunction_Free_AsymptoteOrigin}
{\cal F}_{k,\ell}^{\rm free}(r)
\ \underset{r\to 0}{\sim} \
\sqrt{\sym_\ell}
\, \im^\ell
\, \frac{(kr)^{\ell + 1}}{(2\ell + 1)!!}
.
\end{align}
The Sommerfeld factor is defined as the ratio
\begin{align}
\label{eq:SommerfeldFactors_Def}
S_{\ell}(k)
\equiv
\lim_{r\to 0}
\abs{
\frac{{\cal F}_{k,\ell}(r)}{{\cal F}_{k,\ell}^{\rm free}(r)}
}^2
=
\dfrac{1}{\sym_\ell \, k^{2\ell}}
\abs{
\dfrac{(2\ell+1)!!}{\ell!}
\, \lim_{r\to 0}
\, \dv[\ell]{r}
\left(
\dfrac{{\cal F}_{k,\ell}(r)}{kr}
\right)
}^2
.
\end{align}
The Sommerfeld factors encode the modification of contact-like inelastic processes, such as annihilation of nonrelativistic particles into radiation, due to long-range forces.

\subsection{Irregular solution}
\label{sec:Fundamentals_W0fin}

Besides the physical regular solution ${\cal F}_{k,\ell}(r)$, we will also consider the linearly independent irregular solution ${\cal G}_{k,\ell}(r)$. Following the conventions of Ref.~\cite{Flores:2025qit}, we normalize it consistently with ${\cal F}_{k,\ell}$ such that its large-distance behavior is
\begin{align}
\label{eq:IrregularSolution_AsymptoteInfinity}
{\cal G}_{k,\ell}(r)
\underset{r\to\infty}{\sim}
-
\sqrt{\sym_\ell}\,
\im^\ell e^{\im\theta_\ell(k)}
\cos\qty[
kr- \ell\pi/2+\theta_\ell(k)
].
\end{align}
Thus, the regular and irregular solutions carry the same scattering phase shift, with the irregular solution shifted by $-\pi/2$ relative to the regular one. For the application below, we further restrict to the $s$-wave and to potentials that diverge no faster than $1/r$ as $r\to0$ and can be expanded near the origin as
\begin{align}
\label{eq:Potential_LaurentExpansion}
V(r) =
-\frac{\alpha_{-1}}{r}
+ {\cal O}(r^0).
\end{align}
Using the Frobenius method, the irregular $s$-wave solution admits the expansion (see e.g.~\cite{Flores:2025qit})
\begin{align}
\label{eq:IrregularSolution_Origin}
{\cal G}_{k,0}(r)
=
{\cal G}_{k,0}(0)
\qty[
1
+g_0^{(1)}(k)\,r
-2\mu\alpha_{-1}\,r\ln(\mu r)
+{\cal O}(r^2,r^2\ln r)
].
\end{align}
For real $k$, $g_0^{(1)}(k)$ is real. The choice of scale inside the logarithm is conventional: replacing $\mu$ with any other fixed scale shifts $g_0^{(1)}(k)$ by a momentum-independent constant. Consequently, differences of $g_0^{(1)}$ at different momenta are independent of this convention.

\subsubsection*{Projected $s$-wave Green's function}

The Green's function $G_{\vb{k}}(\vb{r},\vb{r}')$ corresponding to the Schr\"odinger \cref{eq:SchroedingerEquation} enters the unitarization of elastic and inelastic scattering through its projection onto the interaction kernel that is being resummed~\cite{Flores:2024sfy,Flores:2025qit}. For an $s$-wave contact interaction, represented by a delta-function potential, this projection probes the Green's function at $r=r'=0$ and is ultraviolet divergent~\cite{Blum:2016nrz,Flores:2025qit}. For the renormalizable case considered here, the divergent contribution is independent of the incoming momentum, so that the finite momentum-dependent part can be isolated by introducing a subtraction momentum $k_\diamond$. Following Ref.~\cite{Blum:2016nrz}, we define the dimensionless
quantity\footnote{
$T(k;k_\diamond)$ is related to $W_0^{\fin}(k;k_\diamond)$ defined in Ref.~\cite{Flores:2025qit} as
\begin{align*}
W_0^{\fin}(k;k_\diamond) = -
\dfrac{4 k \, k^f}{\sym^f m_T^2} 
\, T(k;k_\diamond),    
\end{align*}
where $m_T$ is the total mass of the initial-state particles, while $k^f$ and $\sym^f$ are the CM momentum and symmetry factor of the final state, respectively.} 
%
\begin{align}
\label{eq:T_def}
T (k;k_\diamond) \equiv   
\dfrac{1}{k} \ 
\qty[g_0^{(1)}(k)-g_0^{(1)}(k_\diamond)] .
\end{align}
Using \cref{eq:IrregularSolution_Origin}, the above becomes
\begin{align}
\label{eq:T}
T(k;k_\diamond)
=
\frac{1}{k}
\lim_{r\to0}
\qty[
\dfrac{{\cal G}'_{k,0}(r)}{{\cal G}_{k,0}(r)} -
\dfrac{{\cal G}'_{k_\diamond,0}(r)}{{\cal G}_{k_\diamond,0}(r)}
].
\end{align}

\clearpage

\section{Variable-phase approach}
\label{sec:VPA}

\subsection{Formulation}\label{sec:VPA_Formulation}

We begin by writing the radial Schr\"odinger equation in the form
\begin{align}
\label{eq:Variable_Phase_SE}
\chi''_{k, \ell}(x) +b_\ell(x)\chi_{k,\ell}(x) =\tilde{V}(x)\chi_{k, \ell}(x),
\end{align}
where $x \equiv kr$, $\chi_{k,\ell} (x) \propto u_{k,\ell} (x/k)$ and
\begin{align}
\label{eq:RescaledPotential_def}
\tilde{V}(x) \equiv \frac{2\mu}{k^2} \, V(x/k)
\quad \text{and} \quad
b_\ell(x) \equiv 1-\frac{\ell(\ell+1)}{x^2} .
\end{align}
Primes denote derivatives with respect to $x$. In the absence of the potential $\tilde{V}(x)$, the linearly independent solutions to \cref{eq:Variable_Phase_SE} are the Riccati--Bessel and Riccati--Neumann functions, $\hat{\jmath}_\ell(x)$ and $\hat{n}_\ell(x)$, respectively, related to the spherical Bessel functions $j_\ell(x)$ and $n_\ell(x)$ via
\begin{subequations}
\label{eq:RiccatiBessel}
\label[pluralequation]{eqs:RiccatiBessel}
\begin{align}
\hat{\jmath}_\ell(x) & \equiv x \, j_\ell(x),
\label{eq:RiccatiBessel_j}
\\
\hat{n}_\ell(x) &\equiv x \, n_\ell(x),
\label{eq:RiccatiBessel_n}
\end{align}
\end{subequations}
which satisfy the Wronskian relation
\begin{align}
\label{eq:Wronskian_RiccatiBessel}
\hat{\jmath}_\ell(x)  \hat{n}_\ell'(x) -
\hat{\jmath}_\ell'(x) \hat{n}_\ell(x)
= 1,
\end{align}
and scale at $x\to0$ and $x\to \infty$ as follows,
\begin{subequations}
\label{eq:RiccatiBessel_Asymptotes}
\label[pluralequation]{eqs:RiccatiBessel_Asymptotes}
\begin{align}
\label{eq:RiccatiBessel_AsymptotesOrigin}
x \to 0:\qquad
\hat{\jmath}_{\ell}(x)
&\simeq
\dfrac{x^{\ell+1}}{(2\ell+1)!!},
&
\hat{n}_{\ell}(x)
&\simeq
-\dfrac{(2\ell-1)!!}{x^{\ell}},
\\[1ex]
\label{eq:RiccatiBessel_AsymptotesInfinity}
x \to \infty:\qquad
\hat{\jmath}_{\ell}(x)
&\sim
\sin (x-\ell\pi/2),
&
\hat{n}_{\ell}(x)
&\sim
-\cos(x-\ell\pi/2),
\end{align}
\end{subequations}
where $(-1)!!=1$. 
Following the variable-phase approach~\cite{Calogero:1967var}, we define two coefficient functions $c_{k,\ell}(x)$ and $s_{k,\ell}(x)$ through
\begin{subequations}
\label{eq:ExpansionRicattiBessel}
\label[pluralequation]{eqs:ExpansionRicattiBessel}
\begin{align}
\chi_{k,\ell}(x)
&=
c_{k,\ell}(x)\hat{\jmath}_\ell(x)
-
s_{k,\ell}(x)\hat n_\ell(x),
\label{eq:ExpansionRicattiBessel_chi}
\\
\chi_{k,\ell}'(x)
&=
c_{k,\ell}(x)\hat{\jmath}_\ell'(x)
-
s_{k,\ell}(x)\hat n_\ell'(x).
\label{eq:ExpansionRicattiBessel_chiprime}
\end{align}
\end{subequations}
The nonvanishing Wronskian in \cref{eq:Wronskian_RiccatiBessel} ensures that these relations uniquely determine $c_{k,\ell}$ and $s_{k,\ell}$ for $x>0$. Differentiating \cref{eq:ExpansionRicattiBessel_chi} and comparing with \eqref{eq:ExpansionRicattiBessel_chiprime} gives
\begin{subequations}
\label{eq:VariationOfParameters}
\label[pluralequation]{eqs:VariationOfParameters}
\begin{align}
\label{eq:VariationOfParameters_Constraint}
c_{k,\ell}'(x)\hat{\jmath}_\ell(x) -
s_{k,\ell}'(x)\hat n_\ell(x)
= 0.
\end{align}
Differentiating \eqref{eq:ExpansionRicattiBessel_chiprime}, and using \cref{eq:Variable_Phase_SE} together with its free version for $\hat{\jmath}_\ell(x)$ and $\hat{n}_\ell(x)$, gives
\begin{align}
\label{eq:VariationOfParameters_Auxiliary}
c_{k,\ell}'(x)\hat{\jmath}_\ell'(x) -
s_{k,\ell}'(x)\hat n_\ell'(x)
=
\tilde{V}(x)\chi_{k,\ell}(x).
\end{align}
\end{subequations}
Solving \cref{eqs:VariationOfParameters} using \eqref{eq:Wronskian_RiccatiBessel}, we obtain
\begin{subequations}
\label{eq:SineCosineFunctionODEs}
\label[pluralequation]{eqs:SineCosineFunctionODEs}
\begin{align}
c_{k,\ell}'(x)
&=
-\tilde{V}(x) \hat n_\ell(x)
[c_{k,\ell}(x)\hat{\jmath}_\ell(x)
-s_{k,\ell}(x)\hat n_\ell(x)],
\label{eq:SineCosineFunctionODEs_Cosine}
\\
s_{k,\ell}'(x)
&=
-\tilde{V}(x)\hat{\jmath}_\ell(x)
[c_{k,\ell}(x)\hat{\jmath}_\ell(x) 
-s_{k,\ell}(x)\hat n_\ell(x)].
\label{eq:SineCosineFunctionODEs_Sine}
\end{align}
\end{subequations}
In the following, we solve \cref{eqs:SineCosineFunctionODEs} to obtain both the regular and irregular wave functions.

\subsection{Regular solution and phase shift}
We select the regular-family solution, $\chi_{k,\ell}^\rgl$, and fix its normalization by requiring
\begin{equation}
\label{eq:BoundaryConditions_chi_Origin}
\lim_{x\to0}
\dfrac{\chi_{k,\ell}^\rgl (x)}{\hat{\jmath}_\ell(x)}
= 1.
\end{equation}
This implies for the corresponding coefficient functions,
\begin{equation}
\label{eq:BoundaryConditions_cs_Origin}
\lim_{x\to0}c_{k,\ell}^\rgl (x)=1,
\qquad
\lim_{x\to0} s_{k,\ell}^\rgl (x)=0.
\end{equation}
Since the potential and the $\chi_{k,\ell}^\rgl$ normalization at the origin are real, $\chi_{k,\ell}^\rgl$, $c_{k,\ell}^\rgl$, and $s_{k,\ell}^\rgl$ are real. The numerical implementation of these conditions at a finite $x_{\min}>0$ is discussed in \cref{sec:VPA_BoundaryCond}. It is also useful to define
\begin{align}
t_{k,\ell} (x) 
\equiv
\dfrac{s_{k,\ell}^\rgl(x)}{c_{k,\ell}^\rgl(x)}.
\end{align}
For potentials satisfying the condition \eqref{eq:Potential_ShortRangeAssumption}, the bounded asymptotic behavior of the Riccati--Bessel and Riccati--Neumann functions in \cref{eq:RiccatiBessel_AsymptotesInfinity}, together with \cref{eqs:SineCosineFunctionODEs}, implies that $c_{k,\ell}^\rgl(x)$ and $s_{k,\ell}^\rgl(x)$ approach finite constants as $x\to\infty$,
\begin{align}
\label{eq:BoundaryConditions_cs_Infinity}
\lim_{x\to \infty} c_{k,\ell}^\rgl(x) 
\equiv c_{k,\ell;\infty}^\rgl,
\qquad
\lim_{x\to \infty} s_{k,\ell}^\rgl(x) 
\equiv s_{k,\ell;\infty}^\rgl ,   
\end{align}
thus
\begin{align}
\label{eq:BoundaryConditions_chi_Infinity}
\chi_{k,\ell}^\rgl (x)    
\ \underset{x\to \infty}{\sim} \
c_{k,\ell;\infty}^\rgl \, \sin(x - \ell\pi/2) + 
s_{k,\ell;\infty}^\rgl \, \cos(x - \ell\pi/2) .
\end{align}
Comparing with the large-distance asymptotic behavior of \cref{eq:PhysicalSolution_AsymptoteInfinity}, we deduce the phase shift through 
\begin{align}
\label{eq:PhaseShift_Tan_Def}
\tan\theta_\ell (k)
\equiv
\lim_{x\to\infty}	
t_{k,\ell} (x) = 
\lim_{x\to\infty}	
\dfrac{s_{k,\ell}^\rgl(x)}{c_{k,\ell}^\rgl(x)}.
\end{align}
The phase shift is only defined modulo $\pi$ through this relation. We resolve this ambiguity by taking $\theta_\ell = \arg(c^{\rgl}_{k,\ell;\infty} + \im\, s^{\rgl}_{k,\ell;\infty}) \in (-\pi,\pi]$, so that $\cos\theta_\ell$ has the sign of $c^{\rgl}_{k,\ell;\infty}$.

\subsubsection*{Standard variable-phase equation}
One could instead directly solve for the phase shift. To do so, we define
\begin{align}
\label{eq:thetabar_def}
\tan\bar{\theta}_{k,\ell} (x)= t_{k,\ell}(x),
\qquad
\lim_{x\to0}\bar{\theta}_{k,\ell}(x)=0,
\end{align}
with the branch chosen continuously from the origin. 
From \cref{eqs:SineCosineFunctionODEs}, we obtain
\begin{subequations}
\begin{align}
t_{k,\ell}'(x) &=
-\tilde{V}(x) \qty[\hat{\jmath}_\ell(x)-t_{k,\ell}(x)\hat{n}_\ell(x)]^2,
\\
\bar{\theta}_{k,\ell}'(x) &=
\dfrac{t_{k,\ell}'(x)}{1+t_{k,\ell}^2(x)}.
\end{align}
\end{subequations}
Combining these equations yields the standard variable-phase equation,
\begin{align}\label{eq:PhaseShift_ODE}
\bar{\theta}_{k,\ell}'(x) =
-\tilde{V}(x)
\qty[
\cos\bar{\theta}_{k,\ell}(x)\hat{\jmath}_\ell(x) -
\sin\bar{\theta}_{k,\ell}(x)\hat{n}_\ell(x)
]^2.
\end{align}
The phase shift is obtained via
\begin{align}\label{eq:PhaseShift_Theta_Def}
	\theta_\ell(k)
	\equiv
    \lim_{x\to\infty}	
	\bar{\theta}_{k,\ell}(x).
\end{align}
Solving directly for the phase shift using \cref{eq:PhaseShift_ODE,eq:PhaseShift_Theta_Def} yields a continuous branch of $\theta_\ell(k)$ that is not restricted to the principal branch of the inverse tangent.  Its low- and high-energy limits may therefore be used to determine the number of bound states through Levinson's theorem, with the usual modification in the presence of an $s$-wave zero-energy resonance~\cite{Calogero:1967var}.
In practice, however, we observe that integrating \cref{eqs:SineCosineFunctionODEs} is numerically more stable, and consequently we use this system to generate the accompanying dataset. Moreover, the functions $c_{k,\ell}^\rgl (x)$ and $s_{k,\ell}^\rgl (x)$ allow us to determine both the phase shift and the Sommerfeld factor, as will be shown below.

\subsubsection{Boundary conditions for the regular solution}
\label{sec:VPA_BoundaryCond}

The exact boundary conditions at the origin are 
given in \cref{eq:BoundaryConditions_cs_Origin}. 
In practice, however, the numerical integration of \cref{eqs:SineCosineFunctionODEs} must begin at a finite $x_{\min}>0$. We therefore derive the leading small-$x$ behavior of $c_{k,\ell}^\rgl (x)$ and $s_{k,\ell}^\rgl (x)$. Substituting the Riccati--Bessel and Riccati--Neumann small-$x$ behavior \eqref{eq:RiccatiBessel_AsymptotesOrigin} into
\cref{eqs:SineCosineFunctionODEs}, and considering the boundary values \eqref{eq:BoundaryConditions_cs_Origin} for $c_{k,\ell}^\rgl$ and $s_{k,\ell}^\rgl$, gives, at leading order,
\begin{subequations}
\begin{align}
\dv{c_{k,\ell}^\rgl (x)}{x}
&\underset{x\to0}{\simeq}
\frac{x}{2\ell+1}\tilde{V}(x),
\\
\dv{s_{k,\ell}^\rgl (x)}{x}
&\underset{x\to0}{\simeq}
-\frac{x^{2\ell+2}}{[(2\ell+1)!!]^2}\tilde{V}(x).
\end{align}
\end{subequations}
Integrating from the origin and using
\cref{eq:BoundaryConditions_cs_Origin}, we obtain
\begin{subequations}
\label{eq:VPS_BCs}
\label[pluralequation]{eqs:VPS_BCs}
\begin{align}
c_{k,\ell}^\rgl (x)
&\underset{x\to0}{\simeq}
1
+
\dfrac{1}{2\ell + 1}
\int_0^x \dd u \, u \,  
\tilde{V}(u),
\\[1ex]
s_{k,\ell}^\rgl (x)
&\underset{x\to0}{\simeq}
-
\dfrac{1}{[(2\ell+1)!!]^2}
\int_0^x \dd u \, u^{2\ell+2} \,
\tilde{V}(u).
\end{align}
\end{subequations}
We use \cref{eqs:VPS_BCs} to initialize the numerical integration of \cref{eqs:SineCosineFunctionODEs} at $x=x_{\min}$. The value of $x_{\min}$ must be sufficiently small for the small-$x$ expansion above to be accurate, while the integration must extend to a sufficiently large $x_{\max}$ for $c_{k,\ell}^\rgl (x)$ and $s_{k,\ell}^\rgl (x)$ to have reached their asymptotic values. The appropriate choices depend on the potential and on the parameters of the problem. In \cref{sec:Yukawa_BCs}, we describe this procedure for a Yukawa potential, although the same general considerations apply to other potentials.

\subsubsection{Sommerfeld factors from the regular solution} 
\label{sec:VPA_SommerfeldFactors}

The variable-phase approach also enables us to determine the Sommerfeld factor. The regular-family solution $\chi_{k,\ell}^\rgl  (x)$ is not identical to the physical radial solution ${\cal F}_{k,\ell}(r)$ defined in \cref{sec:Fundamentals}. The normalization of $\chi_{k,\ell}^\rgl$ is fixed at the origin by \cref{eq:BoundaryConditions_chi_Origin}, whereas that of ${\cal F}_{k,\ell}$ is fixed by its large-distance behavior in \cref{eq:PhysicalSolution_AsymptoteInfinity}.\footnote{
In the Jost-function formalism of \cite{Flores:2025qit}, $\chi_{k,\ell}^\rgl(x)$ is related to the regular solution $\varphi_{k,\ell}(r)$ by $\chi_{k,\ell}^\rgl(x)=k^{\ell+1}\varphi_{k,\ell}(x/k)$.}
Since ${\cal F}_{k,\ell}$ and $\chi_{k,\ell}^\rgl$ are both regular at the origin, they are proportional. Comparing their large-distance behaviors in \cref{eq:PhysicalSolution_AsymptoteInfinity,eq:BoundaryConditions_chi_Infinity}, we obtain
\begin{align}
{\cal F}_{k,\ell}(r)
=
\dfrac
{\sqrt{\sym_\ell}\,\im^\ell e^{\im\theta_\ell(k)}}
{\sqrt{
(c_{k,\ell;\infty}^\rgl)^2 + 
(s_{k,\ell;\infty}^\rgl)^2}
}
\, \chi_{k,\ell}^\rgl (kr).
\end{align}
Using the definition \eqref{eq:SommerfeldFactors_Def} of the Sommerfeld factors and the free solution given in \cref{eq:RadialWaveFunction_Free}, we then find
\begin{align}
S_\ell(k) &=
\lim_{r\to0}
\abs{\frac
{{\cal F}_{k,\ell}(r)}
{{\cal F}^{\rm free}_{k,\ell}(r)}
}^2
=
\dfrac{1}{(c_{k,\ell;\infty}^\rgl)^2 + (s_{k,\ell;\infty}^\rgl)^2}
\ \lim_{x\to0} \abs{
\dfrac{\chi_{k,\ell}^\rgl (x)}{\hat{\jmath}_\ell(x)}
}^2.
\end{align}
Using the normalization of $\chi_{k,\ell}^\rgl$ at the origin, \cref{eq:BoundaryConditions_chi_Origin}, this yields
\begin{align}
\label{eq:VPS_Sommerfeld}
S_\ell(k)
=
\qty[(c_{k,\ell;\infty}^\rgl)^2 + (s_{k,\ell;\infty}^\rgl)^2]^{-1}.
\end{align}
Thus, the functions $c_{k,\ell}^\rgl$ and $s_{k,\ell}^\rgl$, obtained from the integration of \cref{eqs:SineCosineFunctionODEs}, determine both the phase shift and the Sommerfeld factor.

\subsection{Irregular solution and projected $s$-wave Green's function}
\label{sec:VPA_W0fin}

We now return to \cref{eqs:SineCosineFunctionODEs} to construct the irregular solution, $\chi_{k,\ell}^\irr(x)$, from its coefficient functions, $c_{k,\ell}^\irr(x)$ and $s_{k,\ell}^\irr(x)$. The boundary conditions for the irregular solution are imposed at large $x$.  According to \cref{eq:IrregularSolution_AsymptoteInfinity}, the irregular solution is shifted by $-\pi/2$ relative to the regular solution. We therefore choose
\begin{align}
\label{eq:Irregular_cs_Infinity}
c_{k,\ell;\infty}^\irr
=
s_{k,\ell;\infty}^{\rgl},
\qquad
s_{k,\ell;\infty}^\irr
=
-c_{k,\ell;\infty}^{\rgl},
\end{align}
which gives
\begin{align}
\chi_{k,\ell}^\irr(x)
\underset{x\to\infty}{\sim}
-\dfrac{1}{\sqrt{S_\ell(k)}}
\cos\qty[x-\ell\pi/2+\theta_\ell(k)].
\end{align}
Comparison with \cref{eq:IrregularSolution_AsymptoteInfinity} then gives
\begin{align}
\label{eq:PhysicalIrregular_chi}
{\cal G}_{k,\ell}(r)
=
\sqrt{\sym_\ell}\,
\im^\ell e^{\im\theta_\ell(k)}
\sqrt{S_\ell(k)}\,
\chi_{k,\ell}^\irr(kr).
\end{align}
Numerically, we first integrate the regular solution outward to $x_{\max}$ and determine $c_{k,\ell}^{\rgl}(x_{\max})$ and $s_{k,\ell}^{\rgl}(x_{\max})$. Once these have reached their asymptotic values to the desired accuracy, we initialize the irregular solution with
\begin{align}
\label{eq:Irregular_cs_xmax}
c_{k,\ell}^\irr(x_{\max}) =
s_{k,\ell}^{\rgl}(x_{\max})
\quad \text{and} \quad
s_{k,\ell}^\irr(x_{\max}) =
-c_{k,\ell}^{\rgl}(x_{\max}),
\end{align}
and integrate \cref{eqs:SineCosineFunctionODEs} inward.

\subsubsection*{Finite part of the projected $s$-wave Green's function}

For the computation of $T(k;k_\diamond)$, we now specialize to $\ell=0$. Since $x=kr$ and ${\cal G}_{k,0}(r)\propto\chi_{k,0}^\irr(kr)$, their logarithmic derivatives are related by
\begin{align}
\label{eq:Irregular_LogDerivative}
\dfrac{1}{{\cal G}_{k,0}(r)}
\dfrac{\dd {\cal G}_{k,0}(r)}{\dd r}
=
k\,
\dfrac{{\chi_{k,0}^\irr}'(kr)}
{\chi_{k,0}^\irr(kr)},
\end{align}
where the prime on $\chi_{k,0}^\irr$ denotes differentiation with respect to $x$. For the $s$ wave, $\hat{\jmath}_0(x)=\sin x$ and $\hat n_0(x)=-\cos x$, such that
\begin{align}
\label{eq:Irregular_LogDerivative_cs}
\dfrac
{{\chi_{k,0}^\irr}'(x)}
{\chi_{k,0}^\irr(x)}
=
\dfrac
{c_{k,0}^\irr(x)\cos x
-s_{k,0}^\irr(x)\sin x}
{c_{k,0}^\irr(x)\sin x
+s_{k,0}^\irr(x)\cos x}.
\end{align}
Substitution into \cref{eq:T} then yields
\begin{align}
\label{eq:T_VPA}
T(k;k_\diamond)
=
\lim_{r\to0}
\qty[
\dfrac{{\chi_{k,0}^\irr}'(kr)}
{\chi_{k,0}^\irr(kr)}
-
\dfrac{k_\diamond}{k}\,
\dfrac{{\chi_{k_\diamond,0}^\irr}'(k_\diamond r)}
{\chi_{k_\diamond,0}^\irr(k_\diamond r)}
].
\end{align}
If \cref{eq:T_VPA} is evaluated directly, the two logarithmic derivatives must be taken at the same physical radius $r$, so that their momentum-independent divergent parts, given by \cref{eq:IrregularSolution_Origin}, cancel. Alternatively, since the singular behavior at the origin is fixed by \cref{eq:Potential_LaurentExpansion}, the divergent term can be subtracted analytically and $g_0^{(1)}$, up to a momentum-independent constant, can be extracted separately at $k$ and $k_\diamond$. We follow this approach in \cref{sec:ComputingT}.

\subsection{Domain of integration}
\label{sec:VPA_DomainOfIntegration}

The numerical implementation of the variable-phase method requires choosing, for each momentum and partial wave, a finite integration domain $x\in[x_{\min},x_{\max}]$. For the regular solution, $x_{\min}$ must be sufficiently small that the small-$x$ expressions in \cref{eqs:VPS_BCs} provide accurate initial conditions for $c_{k,\ell}^\rgl$ and $s_{k,\ell}^\rgl$. Conversely, $x_{\max}$ must be sufficiently large that the remaining potential tail produces a negligible change in their asymptotic values, and hence in the phase shift and Sommerfeld factor given by \cref{eq:PhaseShift_Tan_Def,eq:VPS_Sommerfeld}. For the irregular solution, the same large-$x$ requirement ensures the accuracy of the boundary conditions in \cref{eq:Irregular_cs_xmax}. For $T(k;k_\diamond)$, the irregular solution is integrated inward separately at $k$ and $k_\diamond$, each down to its own $x_{\min}$.

The appropriate choices of $x_{\min}$ and $x_{\max}$ depend on the form and parameters of the potential $\tilde V(x)$ and on the partial wave. To develop a concrete prescription for selecting the integration domain, we now specialize to the Yukawa potential. The same general considerations can be adapted to other potentials.

\clearpage

\section{The Yukawa potential}\label{sec:Yukawa}

Consider the Yukawa potential,
\begin{align}
\label{eq:YukawaPotential}
V^Y (r)=-\frac{\alpha }{r}e^{-m_{\med}r},
\end{align}
where $\alpha$ may be positive or negative. Alongside the change of variables, $x = kr$, we define two dimensionless parameters,
\begin{align}
\label{eq:DimlessParameters_Yukawa}
\zeta \equiv 
\dfrac{\mu\alpha}{k} =
\dfrac{\alpha}{v_{\rm rel}},
\qquad
\xi \equiv \dfrac{\mu |\alpha|}{m_{\med}},
\end{align}
in terms of which the rescaled potential of \cref{eq:RescaledPotential_def} is
\begin{align}
\label{eq:YukawaPotential_NonDim}
\tilde{V}^Y (x) = 
-\dfrac{2\zeta}{x}
\, e^{-x \, |\zeta|/ \xi}
\ .
\end{align}
The analytic Coulomb limit and its connection with the Yukawa solution are discussed in \cref{App:Coulomb}.

\subsection{Boundary conditions}\label{sec:Yukawa_BCs}

From \cref{eqs:VPS_BCs}, setting $\exp(-x \, |\zeta|/ \xi) \simeq 1$ at $x\to 0$, the limiting behavior of $c_{k,\ell}^\rgl (x)$ and $s_{k,\ell}^\rgl (x)$ at the origin is
\begin{subequations}
\label{eq:VPS_BCs_Yukawa}
\label[pluralequation]{eqs:VPS_BCs_Yukawa}
\begin{align}
c_{k,\ell}^\rgl (x)
&\underset{x\to0}{\simeq}
1
+
\dfrac{1}{2\ell + 1}
\int_0^x \dd u \, u \,
\qty(-\frac{2\zeta}{u})
\simeq 
1 - \frac{2\zeta}{2\ell + 1}x,
\\[1ex]
s_{k,\ell}^\rgl (x)
&\underset{x\to0}{\simeq}
-
\frac{1}{[(2\ell+1)!!]^2}
\int_0^x \dd u \, u^{2\ell+2} \,
\qty(-\frac{2\zeta}{u})
\simeq 
\dfrac{1}{[(2\ell+1)!!]^2}
\dfrac{2\zeta}{(2\ell+2)}
x^{2\ell+2}.
\end{align}
\end{subequations}
These equations establish the numerical starting point for evaluating \cref{eqs:SineCosineFunctionODEs}. Doing so correctly requires careful selection of the initial, nonzero integration point $x_{\min}$. 

\subsubsection*{Determining $x_{\min}$}

The selected value of $x_{\min}$ must (i) lie within the regime where the small-$x$ boundary conditions in \cref{eqs:VPS_BCs_Yukawa} are valid, and (ii) be smaller than the characteristic scales associated with the pairwise balances among the three terms multiplying the wave function in \cref{eq:Variable_Phase_SE}, namely the eigenvalue, centrifugal, and potential terms, which are proportional to $1$, $\ell(\ell+1)/x^2$, and $\tilde{V}(x)$, respectively.

Starting with the small-$x$ approximations, the asymptotic forms of the Riccati--Bessel and Riccati--Neumann functions in \cref{eq:RiccatiBessel_AsymptotesOrigin} require
\begin{subequations}
\label{eq:Yukawa_xmin_Scales}
\label[pluralequation]{eqs:Yukawa_xmin_Scales}
\begin{align}
x \ll 1 .
\label{eq:Yukawa_xmin_Scale0}
\end{align}
In deriving \cref{eqs:VPS_BCs_Yukawa} the Yukawa potential has been approximated by its Coulomb-like short-distance behavior, $e^{-x\,|\zeta|/\xi} \simeq 1$. This requires
$x \, |\zeta|/\xi \ll 1$ and introduces the scale
\begin{align}
x_1(\zeta,\xi) \equiv \xi / |\zeta|.
\label{eq:Yukawa_xmin_Scale1}
\end{align}
Finally, the derivation of \cref{eqs:VPS_BCs} assumes $\chi_{k,\ell}(x)\simeq\hat{\jmath}_\ell(x)$ near the origin. As a self-consistency check, using \cref{eq:ExpansionRicattiBessel_chi,eq:RiccatiBessel_AsymptotesOrigin,eq:VPS_BCs_Yukawa}, we find
\begin{align*}
\frac{\chi_{k,\ell}(x)}{\hat{\jmath}_\ell(x)}
&=
c_{k,\ell}(x) -
s_{k,\ell}(x)
\dfrac{\hat n_\ell(x)}{\hat{\jmath}_\ell(x)}
=
1-\dfrac{\zeta}{\ell+1}x + {\cal O}(x^2).
\end{align*}
The validity of this approximation requires $x \, |\zeta|/(\ell+1) \ll 1$, which defines the scale
\begin{align}
x_2 (\ell,\zeta) \equiv
(\ell+1)/\abs{\zeta}.
\label{eq:Yukawa_xmin_Scale2}
\end{align}
We next consider the pairwise balances among the terms in the Schr\"odinger \cref{eq:Variable_Phase_SE}. The balance in magnitude between the eigenvalue and potential terms occurs at
\begin{align}
x_3 (\zeta,\xi) \equiv \dfrac{\xi}{|\zeta|}
\, {\cal W}_0 \qty(\dfrac{2\zeta^2}{\xi}),
\label{eq:Yukawa_xmin_Scale3}
\end{align}
where ${\cal W}_0(z)$ is the Lambert ${\cal W}$ function. For $\ell \geq 1$, the balance between the centrifugal and potential terms may produce two crossings. The first crossing encountered from the origin, which is the relevant one for determining $x_{\min}$, introduces the scale
\begin{align}
\label{eq:Yukawa_xmin_Scale4}
x_4 (\ell,\zeta,\xi) \equiv
\begin{cases}
\displaystyle
\frac{\xi}{|\zeta|}
\left[
-{\cal W}_{0}\qty(-\dfrac{\ell(\ell+1)}{2\xi})
\right],
&
\xi \geq \dfrac{e}{2} \, \ell(\ell+1),
\\[2ex]
+\infty,
&
\xi < \dfrac{e}{2} \, \ell(\ell+1).
\end{cases}
\end{align}
In the second case, the magnitude of the centrifugal term remains larger than that of the potential term for all $x$, so this comparison imposes no additional restriction on $x_{\min}$.
The balance between the eigenvalue and centrifugal terms occurs at 
\begin{align}
x_5 (\ell)=\sqrt{\ell(\ell+1)} .  
\label{eq:Yukawa_xmin_Scale5} 
\end{align}
\end{subequations}
For $\ell \geq 1$, this exceeds unity, so it is superseded by the condition in \cref{eq:Yukawa_xmin_Scale0} and need not be included separately.

Collecting \cref{eqs:Yukawa_xmin_Scales}, we set the lower boundary of the integration domain to
\begin{subequations}\label{eq:xmin_numerical}
\begin{align}
x_{\min}(\ell=0,\zeta,\xi)
&=
\varepsilon \cdot
\min \qty{
1,
\ x_1(\zeta,\xi),
\ x_2(0,\zeta),
\ x_3(\zeta,\xi)
},
\\
x_{\min}(\ell \geq 1,\zeta,\xi)
&=
\varepsilon \cdot
\min \qty{
1,
\ x_1(\zeta,\xi),
\ x_2(\ell,\zeta),
\ x_3(\zeta,\xi),
\ x_4(\ell,\zeta,\xi)
},
\end{align}
\end{subequations}
where $\varepsilon \ll 1$. The accompanying code sets $\varepsilon = 10^{-7}$.

\begin{figure}[ht!]
\centering
\includegraphics[width=0.85\linewidth]{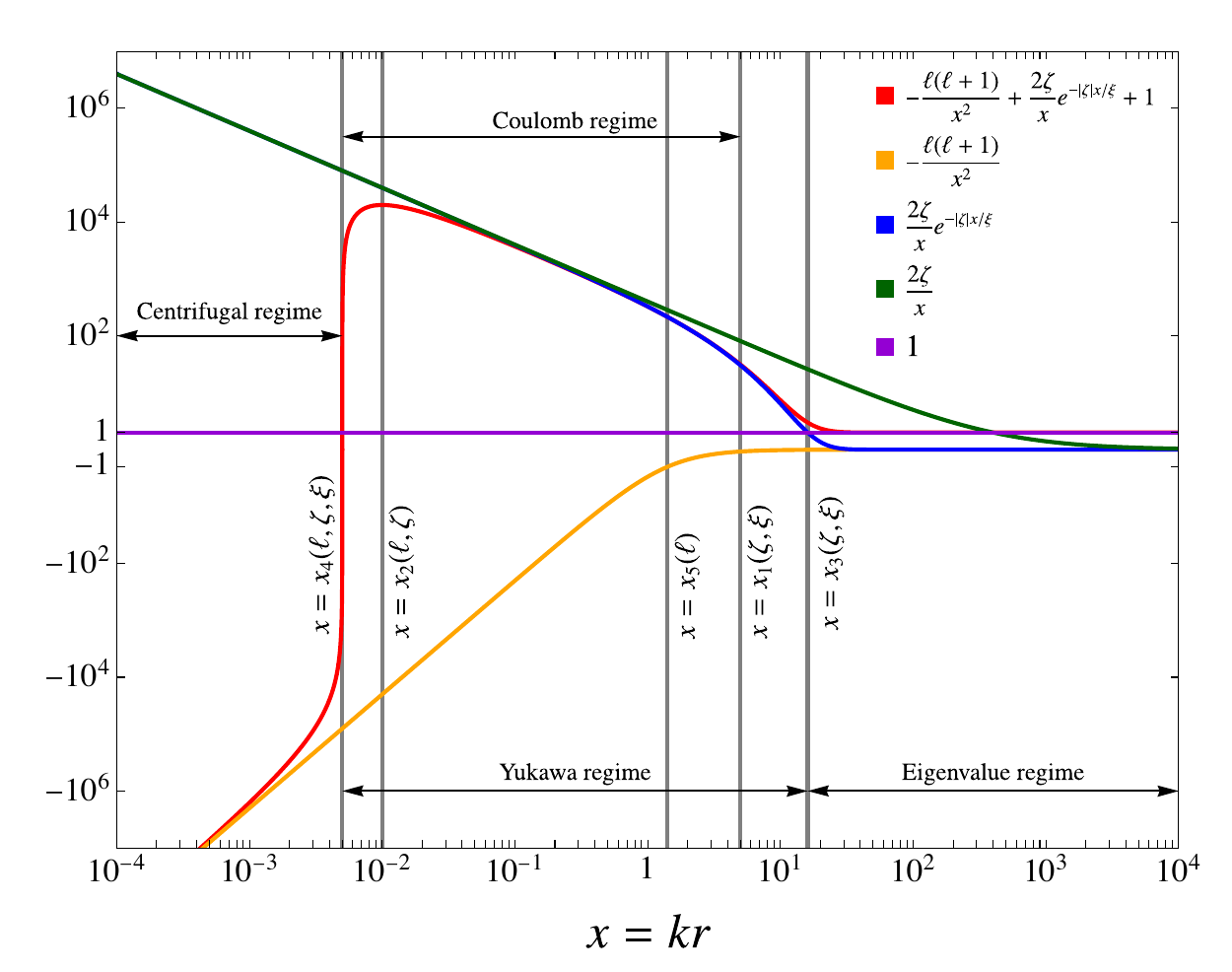}
\caption{Illustration of the different length scales needed to compute the numerical domain of integration. Here we examine an attractive Yukawa potential, \cref{eq:YukawaPotential_NonDim}, with $\ell=1$, $\zeta=200$ and $\xi=10^{3}$.}
\label{fig:ComparisonOfTerms}
\end{figure}

\subsubsection*{Determining $x_{\max}$}

At sufficiently large distances, the eigenvalue term dominates over both the centrifugal and potential contributions. The balance with the centrifugal term occurs at the scale defined in \cref{eq:Yukawa_xmin_Scale5}, while the balance with the potential term occurs at the scale defined in \cref{eq:Yukawa_xmin_Scale3}. We also take the endpoint to lie well beyond the screening scale in \cref{eq:Yukawa_xmin_Scale1}, so that the Yukawa potential is exponentially suppressed. We therefore set
\begin{equation}\label{eq:xmax_numerical}
x_{\max}(\ell,\zeta,\xi)
=
\Lambda
\cdot
\max
\qty{
1,
x_1(\zeta,\xi),
x_3(\zeta,\xi),
x_5(\ell)
},
\end{equation}
where $\Lambda\gg1$. In the accompanying code, we take $\Lambda=10^4$. The same value of $x_{\max}$ is used to initialize the inward integration of the irregular solution.

\subsection{Regular solution: phase shifts and Sommerfeld factors}\label{sec:Yukawa_Regular}

\begin{figure}[t!]
    \centering
    \includegraphics[width=0.47\linewidth]{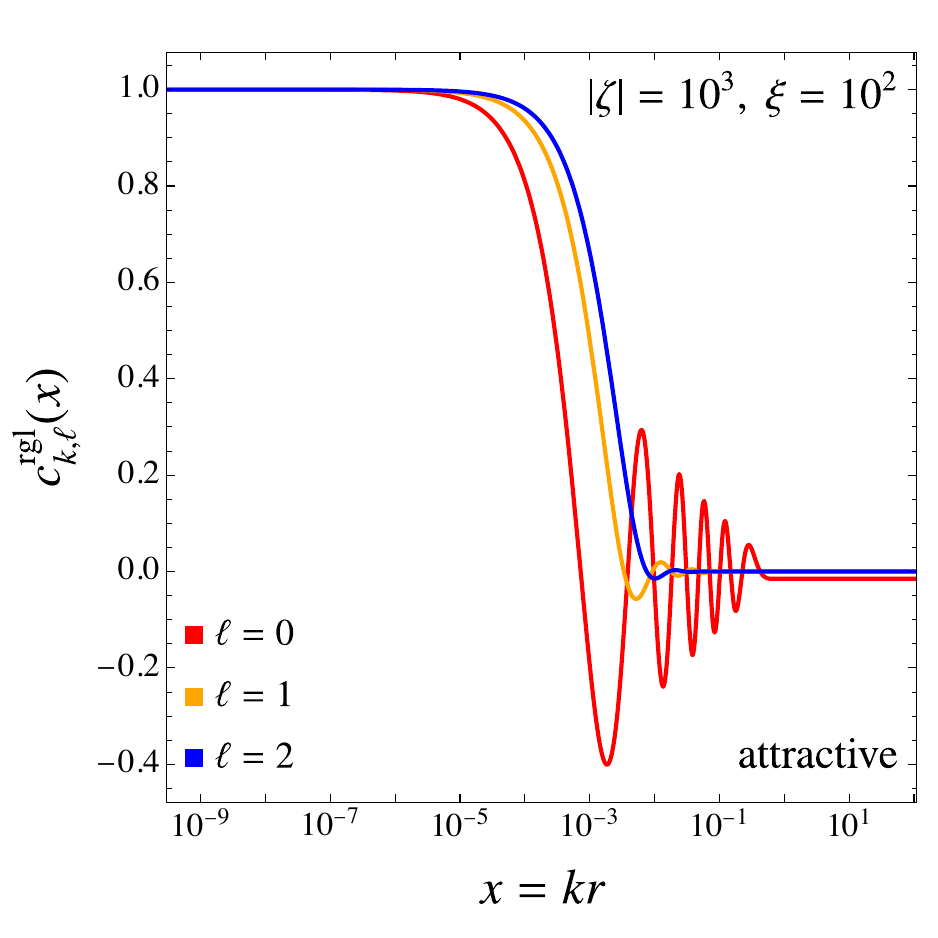}
    \includegraphics[width=0.47\linewidth]{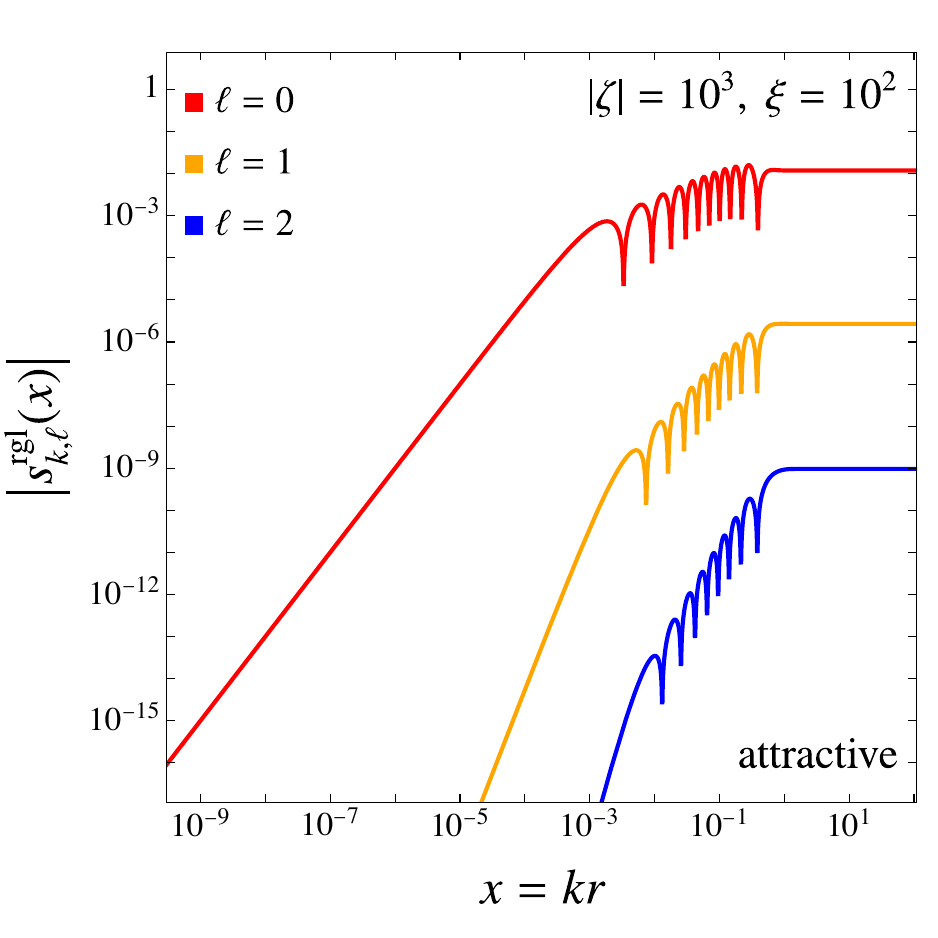}
    \caption{Regular coefficient functions $c_{k,\ell}^\rgl(x), s_{k,\ell}^\rgl(x)$ for an attractive Yukawa potential with $\zeta = 10^3$, $\xi = 10^2$ and for several values of $\ell$.}
    \label{fig:cl_sl}
\end{figure}

As a first illustration of the numerical solution, \cref{fig:cl_sl} shows the evolution of $c_{k,\ell}^\rgl(x)$ and $s_{k,\ell}^\rgl(x)$ for the Yukawa potential. After an initial period of slow evolution followed by oscillations, both functions approach asymptotic values, from which the phase shift and Sommerfeld factor can be determined via \cref{eq:PhaseShift_Tan_Def,eq:VPS_Sommerfeld}, respectively.

\begin{figure}[t!]
    \centering
    \includegraphics[width=0.45\linewidth]{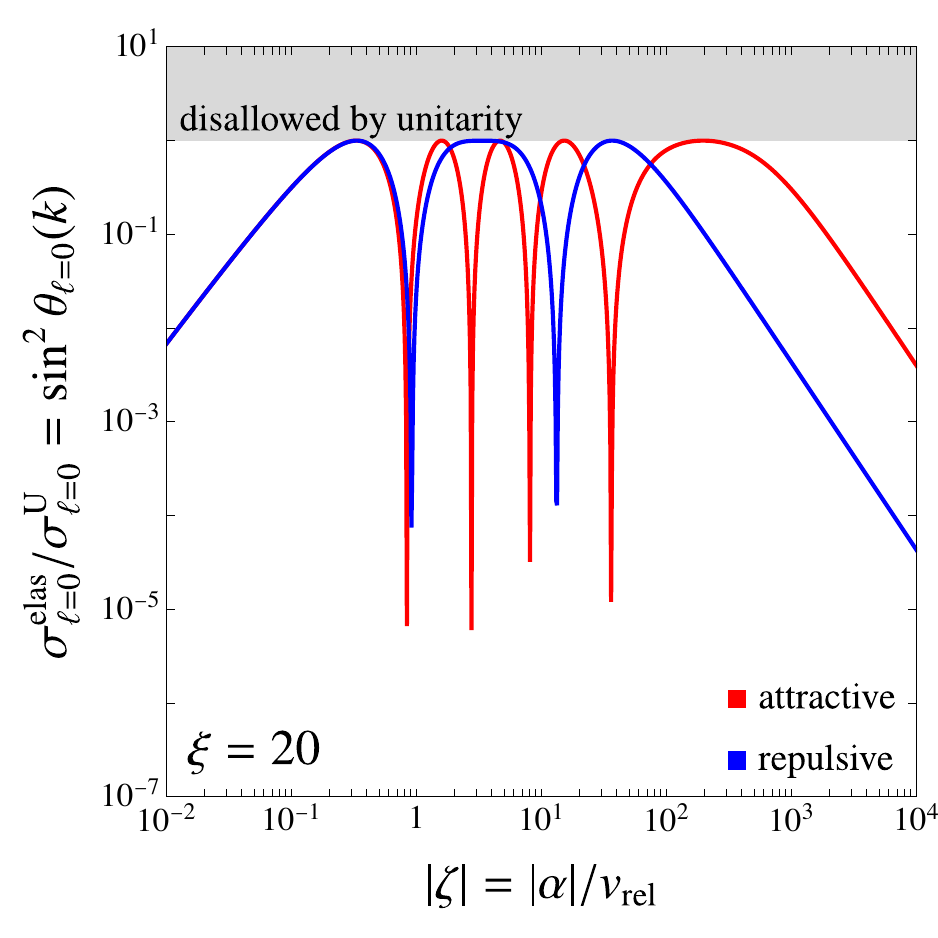}
    \includegraphics[width=0.45\linewidth]{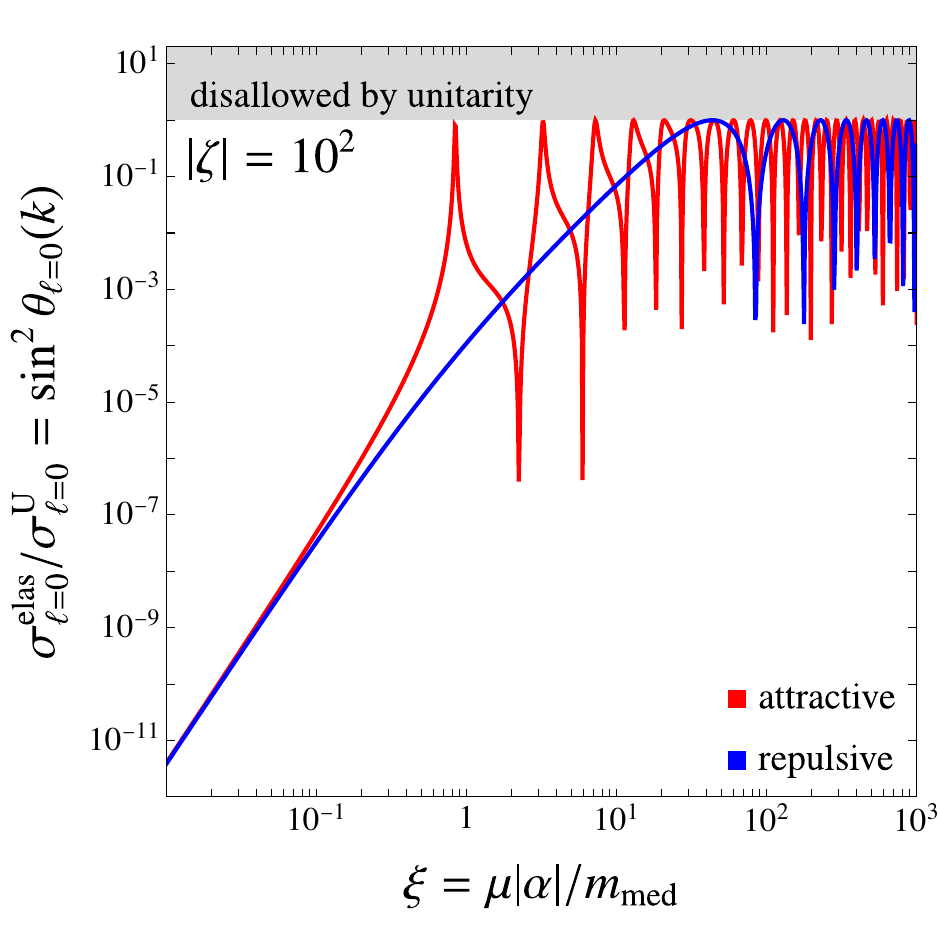}
    \caption{The $s$-wave elastic cross section normalized to its unitarity limit, for attractive and repulsive Yukawa potentials, as a function of $|\zeta|$ (\textit{left}) and $\xi$ (\textit{right}).}
    \label{fig:elastic_CS_l=0}
\end{figure}

\Cref{fig:elastic_CS_l=0} shows the $\ell=0$ elastic cross section for both a repulsive and an attractive potential. Rather than displaying the phase shift itself, which is defined only modulo $\pi$, we plot the normalized quantity $\sin^2\theta_{\ell = 0}$, which is unambiguously defined and proportional to the elastic cross section.

\begin{figure}[t!]
    \centering
    \includegraphics[width=0.45\linewidth]{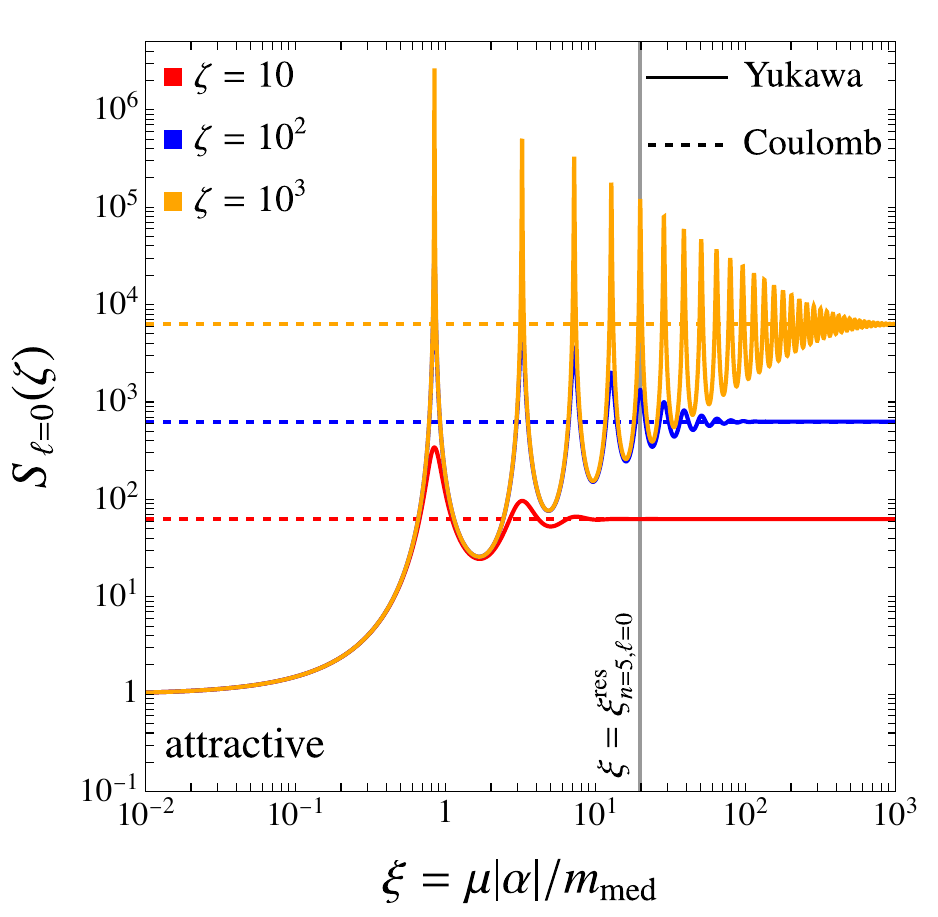}
    \includegraphics[width=0.45\linewidth]{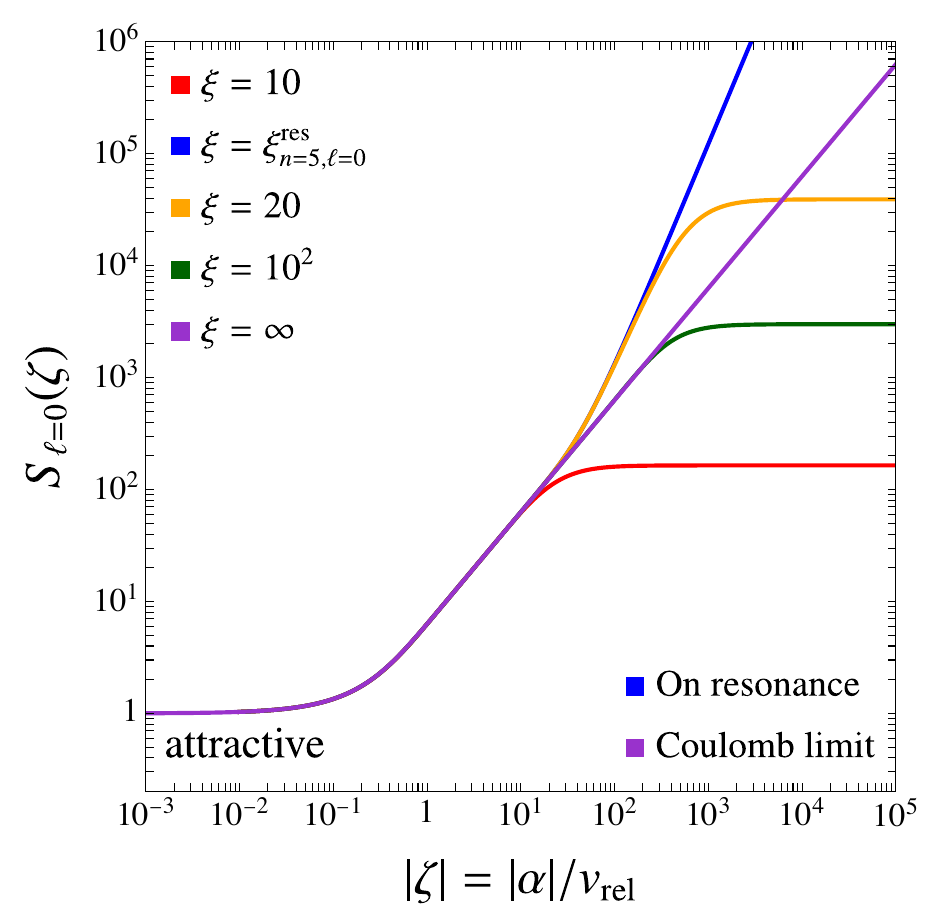}
    \caption{The $s$-wave Sommerfeld factor for various combinations of parameters, assuming an attractive Yukawa potential. Here, $\xi = \xi_{n=5,\ell = 0}^{\rm res}$ 
    corresponds to the fifth resonance, labeled in the left panel. Both panels are compared to the analytic result for the Coulomb potential,~\cref{eq:Coulomb_Sommerfeld}.}
    \label{fig:Sommerfeld_l=0}
\end{figure}

\Cref{fig:Sommerfeld_l=0} illustrates the $\ell = 0$ Sommerfeld factor for an attractive potential. The left panel shows the emergence of resonances as $\xi$ increases. In the limit $\xi\to\infty$ at fixed $\zeta$, the Sommerfeld factor converges to the analytic Coulomb result given in~\cref{eq:Coulomb_Sommerfeld}. The right panel instead shows the Sommerfeld factor as a function of $\zeta$ for fixed values of $\xi$. Away from resonances, the enhancement saturates at large $\zeta$ due to the finite range of the Yukawa interaction. At the bound-state thresholds, however, it grows as $S_0\propto\zeta^2$, without saturation. This differs from the Coulomb behavior, $S_0^C\propto\zeta$ at $\zeta\gg1$, which also exhibits unbounded growth but with a milder scaling.

\begin{figure}[t!]
    \centering
    \includegraphics[width=0.9\linewidth]{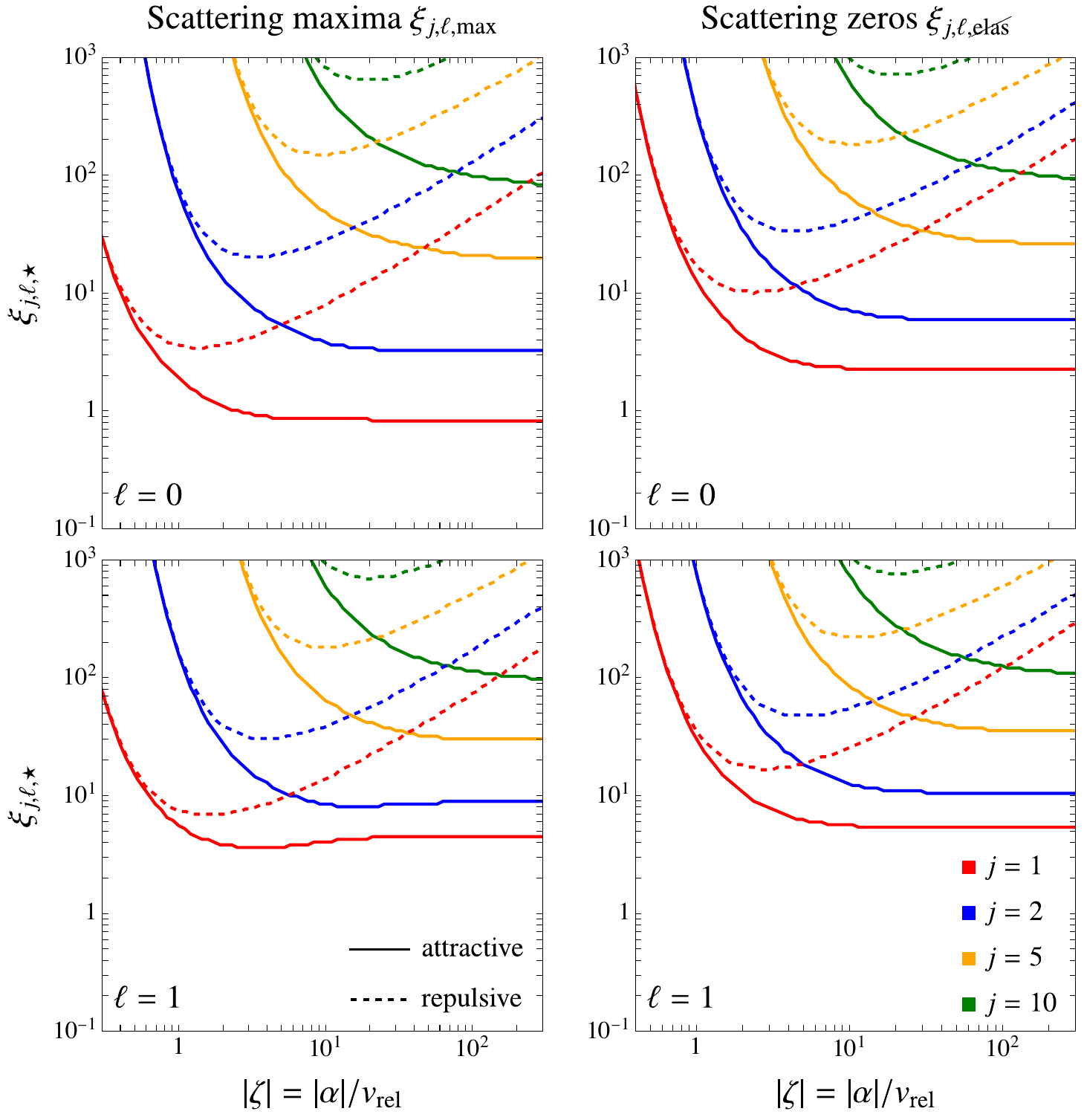}
    \caption{Trajectories of elastic-scattering extrema for attractive (solid) and repulsive (dashed) Yukawa potentials, showing maxima (\emph{left}) and zeros (\emph{right}) for $\ell=0$ (\emph{upper}) and $\ell=1$ (\emph{lower}). Maxima satisfy $\theta_\ell=\pi/2\pmod{\pi}$ and saturate the partial-wave unitarity bound, while zeros satisfy $\theta_\ell=0\pmod{\pi}$. For attractive potentials, the maxima approach the bound-state thresholds and the zeros approach the antiresonance positions as $\zeta\to\infty$. For $\ell=1$ and low $n$, quasi-bound states produce scattering maxima at $\xi<\xi_{n,\ell}$, which approach the corresponding thresholds at low velocities; this behavior is not universal for higher $n$. For repulsive potentials, the extrema shift toward larger $\xi$ as $|\zeta|$ increases, with the first maximum marking the onset of quasi-Coulomb oscillations.}
    \label{fig:positions_extrema}
\end{figure}

\bigskip
\noindent
The behavior shown in \cref{fig:elastic_CS_l=0,fig:Sommerfeld_l=0,fig:positions_extrema} can be understood in terms of the familiar Born, resonant, classical, and Coulomb regimes. A useful estimate of the angular modes relevant for elastic scattering is obtained classically by considering impact parameters comparable to the range of the force,
\begin{align}
\ell_{\rm clas}\sim r_{\rm force}\,k
\sim \mu\vrel / m_{\med}
=\xi / |\zeta| .
\label{eq:ell_clas}
\end{align}
With this estimate in mind, we distinguish the following regimes.
\begin{description}

\item[Born regime, $\bm{\xi\ll1}$:]
The interaction is sufficiently weak for perturbation theory to apply. The phase shifts are small, the Sommerfeld factors approach unity, and the partial-wave elastic cross sections remain well below their unitarity bounds. At leading order, attractive and repulsive potentials yield the same elastic cross sections.

\item[Resonant regime for attractive potentials, $\bm{0.84\lesssim\xi\lesssim|\zeta|}$:]
Scattering is typically dominated by the lowest partial waves, $\ell_{\rm clas}\lesssim1$. Bound states appear as $\xi$ increases beyond discrete critical values $\xi_{n,\ell}$, at which the corresponding binding energies vanish; the lowest threshold is $\xi_{1,0} \simeq 0.84$. Near these thresholds, the elastic scattering cross sections and Sommerfeld factors can exhibit strong resonant enhancements.

For the $s$ wave, the resonance occurs when a bound state reaches zero energy. At $\xi=\xi_{n,0}$, the phase shift approaches $\theta_0\to\pi/2\pmod{\pi}$ as $\zeta\to\infty$, and the elastic cross section approaches its unitarity limit.

For higher partial waves, the centrifugal barrier leads to a qualitatively different resonance structure~\cite{Beneke:2024iev}. For low $n$ and $\xi$ slightly below $\xi_{n,\ell}$, a bound-state level may continue into the positive-energy continuum as a quasi-bound state, temporarily trapped behind the centrifugal barrier. Such a state has a finite lifetime and can generate a narrow scattering resonance when its energy matches the kinetic energy of the incoming particles. The corresponding cross-section maximum therefore occurs at a momentum-dependent value of $\xi$, generally below $\xi_{n,\ell}$, and approaches the bound-state threshold as $\zeta\to\infty$.  For $\xi>\xi_{n,\ell}$, the same level becomes a genuine bound state with negative energy, so its associated positive-energy resonance is absent. The resonant behavior is thus asymmetric about the threshold. Moreover, unlike the $s$ wave, higher partial waves do not generally saturate their elastic unitarity bounds in the zero-momentum limit, even at the bound-state thresholds.

Between successive bound-state thresholds, antiresonances can also occur, leading to a strong suppression of low-energy elastic scattering, known as the Ramsauer--Townsend effect.

Repulsive potentials do not support bound-state resonances.

\item[Classical regime, $\bm{|\zeta|\lesssim1\ll\xi}$:]
Here, $\ell_{\rm clas}\gg1$, and many partial waves contribute to scattering. Their cumulative contribution can yield a sizable elastic cross section even when the individual phase shifts are small. 

\item[Coulomb regime, $\bm{1\ll|\zeta|\ll\xi}$:]
The interaction becomes effectively Coulomb-like, and many partial waves contribute ($\ell_{\rm clas}\gg1$). In the strict Coulomb limit, the total elastic cross section diverges, while the finite mediator mass of the Yukawa potential screens the interaction and renders the partial-wave sum finite.

The normalized partial-wave elastic cross sections, $\sigma_\ell^\elas/\sigma_\ell^{\U}=\sin^2\theta_\ell$, oscillate between zero and their unitarity limits as $\xi$ or $\zeta$ varies. Individual Yukawa partial-wave cross sections do not converge pointwise to their Coulomb counterparts as $\xi\to\infty$, although the Sommerfeld factors do, as discussed in \cref{App:Coulomb}.

At fixed $\zeta$, the positions of the scattering maxima and zeros define the trajectories $\xi_{j\ell,\max}(\zeta)$ and $\xi_{j\ell,\cancel{\elas}}(\zeta)$, respectively, where $j$ labels the different branches. These trajectories are illustrated for $\ell=0,1$ in \cref{fig:positions_extrema}. For a repulsive potential, the first scattering maximum as $\xi$ increases at fixed $\zeta$ provides a practical criterion for the onset of Coulomb-like oscillations.

For attractive potentials, as $\zeta\to\infty$, the maxima associated with resonances approach the bound-state thresholds, while the zeros approach the antiresonance positions, where the leading low-energy elastic amplitude vanishes:
\begin{align}
\xi_{n,\ell} = 
\lim_{\zeta\to\infty}\xi_{n\ell;\max}(\zeta)
\qquad \text{and} \qquad
\xi_{j\ell,\cancel{\elas}}^{(\infty)}
\equiv \lim_{\zeta\to\infty}
\xi_{j\ell,\cancel{\elas}}(\zeta).
\label{eq:Resonance+AntiresonanceXiValues}
\end{align}
For repulsive potentials, the extrema instead shift toward increasingly large $\xi$ as $|\zeta|$ grows, without approaching finite limiting values.  Numerically, the positions of the repulsive $s$-wave elastic-scattering zeros approximately scale as $\xi_{j,0;\cancel{\elas}}(\zeta)\propto|\zeta|^{0.84}$ over the range shown in \cref{fig:positions_extrema}.
\end{description}
Both the classical and Coulomb regimes involve many contributing partial waves, for which a semiclassical description is useful. We discuss the WKB approximation in \cref{sec:Yukawa_WKB,app:WKB_Approx}.

\subsection{Irregular solution: $T(k;k_\diamond)$}\label{sec:ComputingT}

\begin{figure}[ht!]
    \centering
    \includegraphics[width=0.49\linewidth]{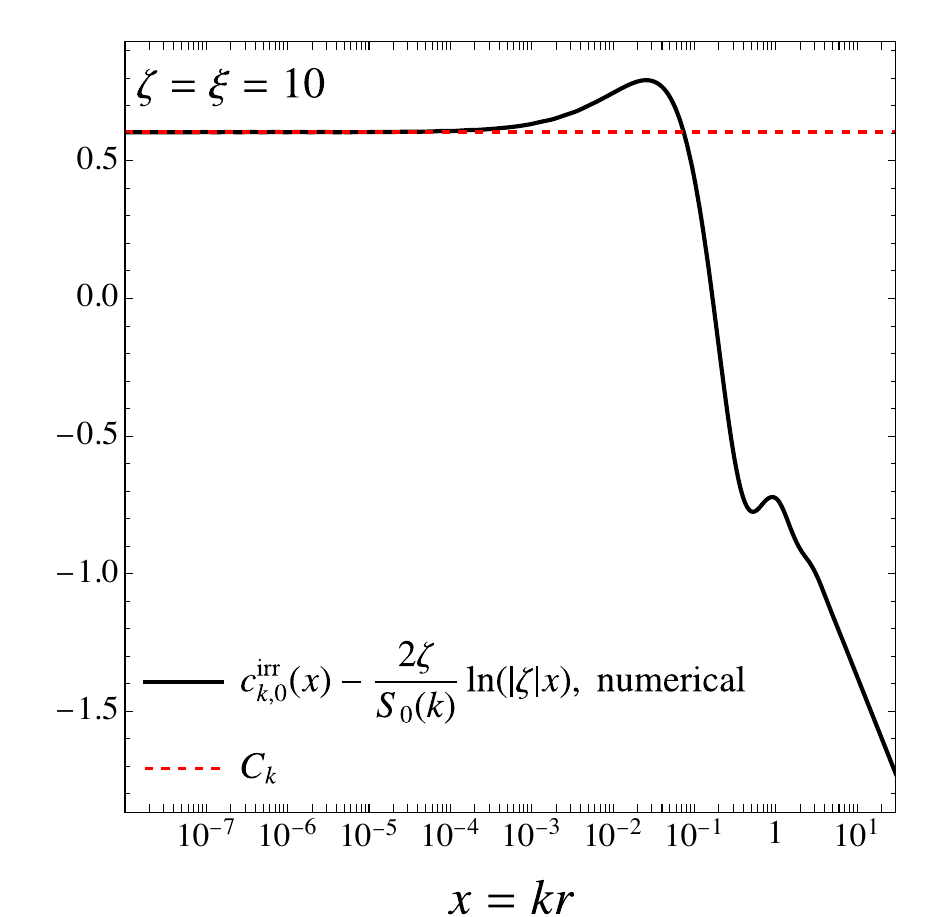}
    \includegraphics[width=0.49\linewidth]{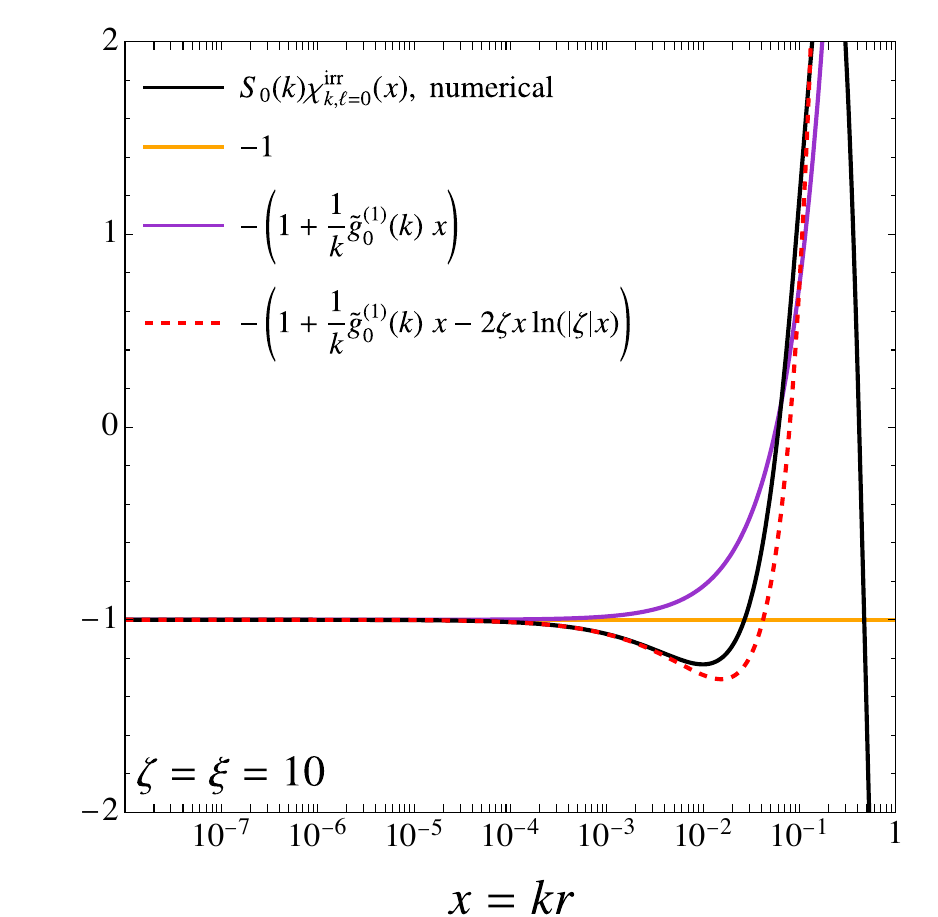}
    \caption{Numerical computation of the irregular solution. The left panel shows $c_{k,0}^{\irr}(x)$ with the logarithmic behavior at the origin subtracted, allowing the linear coefficient of the irregular solution to be determined via~\cref{eq:Tilde_g_Relation}. The right panel shows successive approximations to the irregular solution obtained using the value of $\tilde{g}_{0}^{(1)}(k)/k$ determined from~\cref{eq:Tilde_g_Relation}.}
    \label{fig:Irreg_Approx}
\end{figure}

Having obtained the phase shift and Sommerfeld factor from the regular solution, we can now describe a practical method for computing \cref{eq:T_def}. Since $T(k;k_\diamond)$ involves only the $s$-wave projected Green's function, we recall that the $\ell = 0$ irregular solution can be decomposed as
\begin{align}\label{eq:Irr_Chi_Decompos}
\chi_{k,0}^{\irr}(x)
&=
c_{k,0}^{\irr}(x)\hat{\jmath}_0(x)
-
s_{k,0}^{\irr}(x)\hat n_0(x),
\end{align}
with $x\to\infty$ boundary conditions for $c_{k,0}^{\irr}(x)$ and $s_{k,0}^{\irr}(x)$ given by~\cref{eq:Irregular_cs_Infinity}. To examine the small-$x$ behavior of the above decomposition we note that the Wronskian, $W[\chi^{\rgl}_{k,0}(x),\chi^{\irr}_{k,0}(x)]$, is independent of $x$. As $x\to\infty$,~\cref{eq:Wronskian_RiccatiBessel,eq:Irregular_cs_Infinity} give
\begin{align}
W[\chi^{\rgl}_{k,0},\chi^{\irr}_{k,0}] =(c^{\rgl}_{k,0;\infty})^2+(s^{\rgl}_{k,0;\infty})^2= \frac{1}{S_0(k)}.
\end{align}
As $x\to0$, $\chi^{\rgl}_{k,0}(x)\simeq x$ while, by~\cref{eq:IrregularSolution_Origin},
$\chi^{\irr\,\prime}_{k,0}(x) = \mathcal{O}(\ln x)$, so that
\begin{align}
W[\chi^{\rgl}_{k,0}(x),\chi^{\irr}_{k,0}(x)]
\underset{x\to 0}{\longrightarrow}
-\chi^{\irr}_{k,0}(0),
\end{align}
implying $\chi_{k,0}^{\irr}(x\to 0) = -1/S_0(k)$. Moreover, the relation between $\chi_{k,0}^{\irr}(x)$ and ${\cal G}_{k,0}(r)$ given in~\cref{eq:PhysicalIrregular_chi} allows $\chi_{k,0}^{\irr}(x)$ to be expanded as in~\cref{eq:IrregularSolution_Origin},
\begin{align}\label{eq:Chi_irr_Small_x}
\chi_{k,0}^{\irr}(x)
= 
-\frac{1}{S_0(k)}
\qty[1 + \frac{1}{k}\tilde{g}_{0}^{(1)}(k)\, x - 2\zeta x\, \ln\qty(|\zeta| x) + {\cal O}(x^2,x^2\ln x)]
.
\end{align}
Here, we have adjusted the argument of the logarithmic term to be $|\zeta| x$ rather than the one appearing in~\cref{eq:IrregularSolution_Origin}. This amounts to a redefinition of the coefficient of the linear term. Specifically, the function $\tilde{g}_{0}^{(1)}$ is related to $g_0^{(1)}$, as defined in~\cref{eq:IrregularSolution_Origin}, by
\begin{align}\label{eq:g_tilde_def}
\tilde{g}_{0}^{(1)}(k) = g_{0}^{(1)}(k) + 2\alpha\mu\ln|\alpha|
.
\end{align}
Since $|\zeta|x = \mu|\alpha| r$, the shift in \cref{eq:g_tilde_def} is independent of $k$ and cancels in~\cref{eq:T_def}, so that
\begin{align}\label{eq:T_Eval}
T(k;k_\diamond)=\frac{\tilde{g}^{(1)}_0(k)}{k}-\frac{k_\diamond}{k}\,\frac{\tilde{g}^{(1)}_0(k_\diamond)}{k_\diamond}
.
\end{align}
To determine $\tilde{g}_{0}^{(1)}(k)$, we consider \cref{eqs:SineCosineFunctionODEs} to lowest order in $x$. Using $\tilde V(x)\simeq -2\zeta/x$ together with~\cref{eq:Chi_irr_Small_x}, we find
\begin{subequations}\label{eqs:cs_Irreg_Small_x_ODEs}
\begin{align}
\dv{x} c_{k,0}^{\irr}(x) &\underset{x\to0}{\simeq} \frac{2\zeta}{S_{0}(k)\, x},\\[1ex]
\dv{x} s_{k,0}^{\irr}(x) &\underset{x\to0}{\simeq} -\frac{2\zeta}{S_0(k)}.
\end{align}
\end{subequations}
Integrating \cref{eqs:cs_Irreg_Small_x_ODEs} and using~\cref{eq:Chi_irr_Small_x}, we obtain
\begin{subequations}\label{eqs:cs_Irreg_Small_x_Solutions}
\begin{align}
s_{k,0}^{\irr}(x)
&\underset{x\to0}{\simeq} 
-\frac{1}{S_{0}(k)}\qty[1 + 2\zeta x],\\[1ex]
c_{k,0}^{\irr}(x)
&\underset{x\to0}{\simeq} 
\frac{2\zeta}{S_{0}(k)}\ln\qty(|\zeta| x) + C_{k},
\end{align}
\end{subequations}
where $C_{k}$ is a constant of integration. Inserting~\cref{eqs:cs_Irreg_Small_x_Solutions} into \cref{eq:Irr_Chi_Decompos} and matching to~\cref{eq:Chi_irr_Small_x} order by order in $x$, we find
\begin{align}\label{eq:Tilde_g_Relation}
\frac{1}{k}\tilde{g}_{0}^{(1)}(k) = - S_{0}(k) \qty(C_{k}-\frac{2\zeta}{S_{0}(k)})
.
\end{align}
Having established this relationship, we now describe how to numerically obtain $T(k;k_\diamond)$. Starting from the regular solution of \cref{sec:Yukawa_Regular}, we numerically integrate~\cref{eqs:SineCosineFunctionODEs} inward from $x_{\max}$, defined in~\cref{eq:xmax_numerical}, imposing the boundary conditions given by~\cref{eq:Irregular_cs_xmax}. As before, we set the lower limit of integration using~\cref{eq:xmin_numerical}, evaluated at $k$. Once $c_{k,0}^{\irr}$ has been obtained, we extract the constant of integration according to
\begin{align}
C_k^{\texttt{num}} \equiv c_{k,0}^{\irr}(x_{\min}) - \frac{2\zeta}{S_{0}(k)}\ln\qty(|\zeta| x_{\min})
.
\end{align}
The left panel of \cref{fig:Irreg_Approx} illustrates this procedure. With the constant of integration determined, $\tilde{g}_0^{(1)}(k)/k$ follows immediately from~\cref{eq:Tilde_g_Relation}. This process is then repeated for $k_\diamond$, after which \cref{eq:T_Eval} can be evaluated directly. This procedure is used to generate the accompanying dataset.

\begin{figure}[t!]
    \centering
    \includegraphics[width=0.45\linewidth]{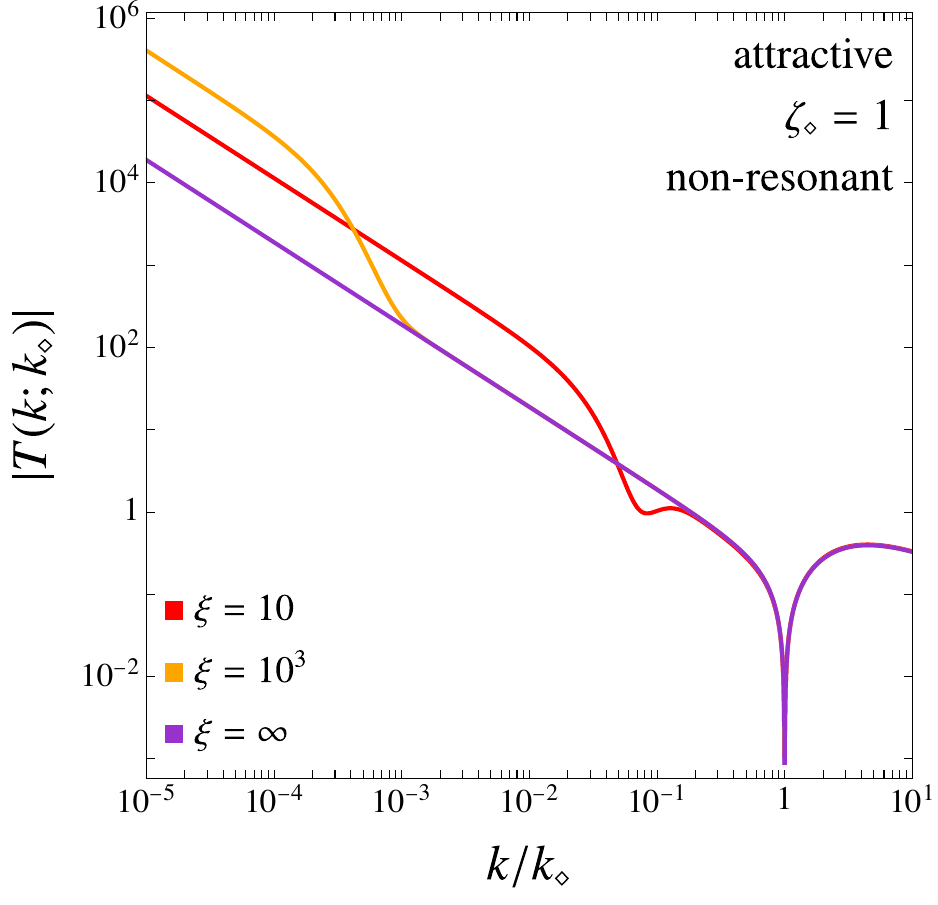}
    \includegraphics[width=0.45\linewidth]{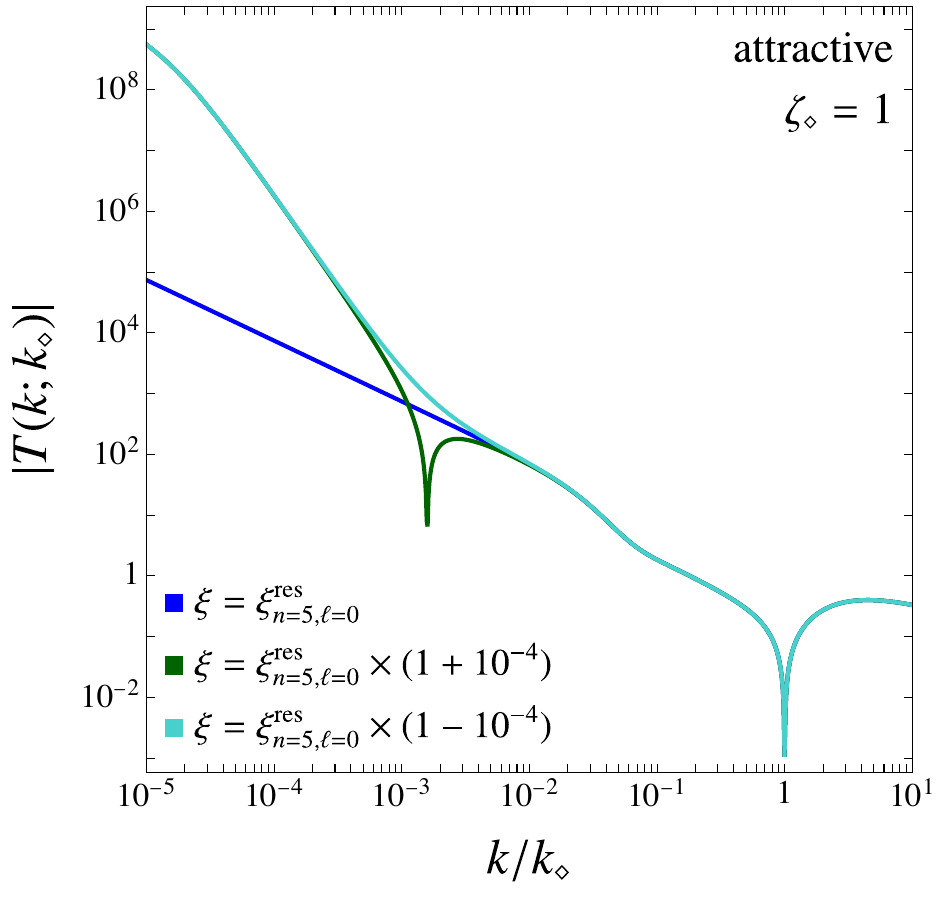}
    \caption{
    (\emph{left}) $|T(k;k_\diamond)|$ as a function of $k/k_\diamond$ for an attractive Yukawa potential with $\zeta_\diamond = 1$. The numerical result follows the Coulomb limit of~\cref{eq:Coulomb_T} for $k/k_\diamond\gtrsim 1/\xi$.  
    (\emph{right}) Sensitivity of $T(k;k_\diamond)$ around bound-state thresholds.    
    }
\label{fig:T_Yukawa_vs_k}
\end{figure}

\begin{figure}[t!]
    \centering    \includegraphics[width=0.725\linewidth]{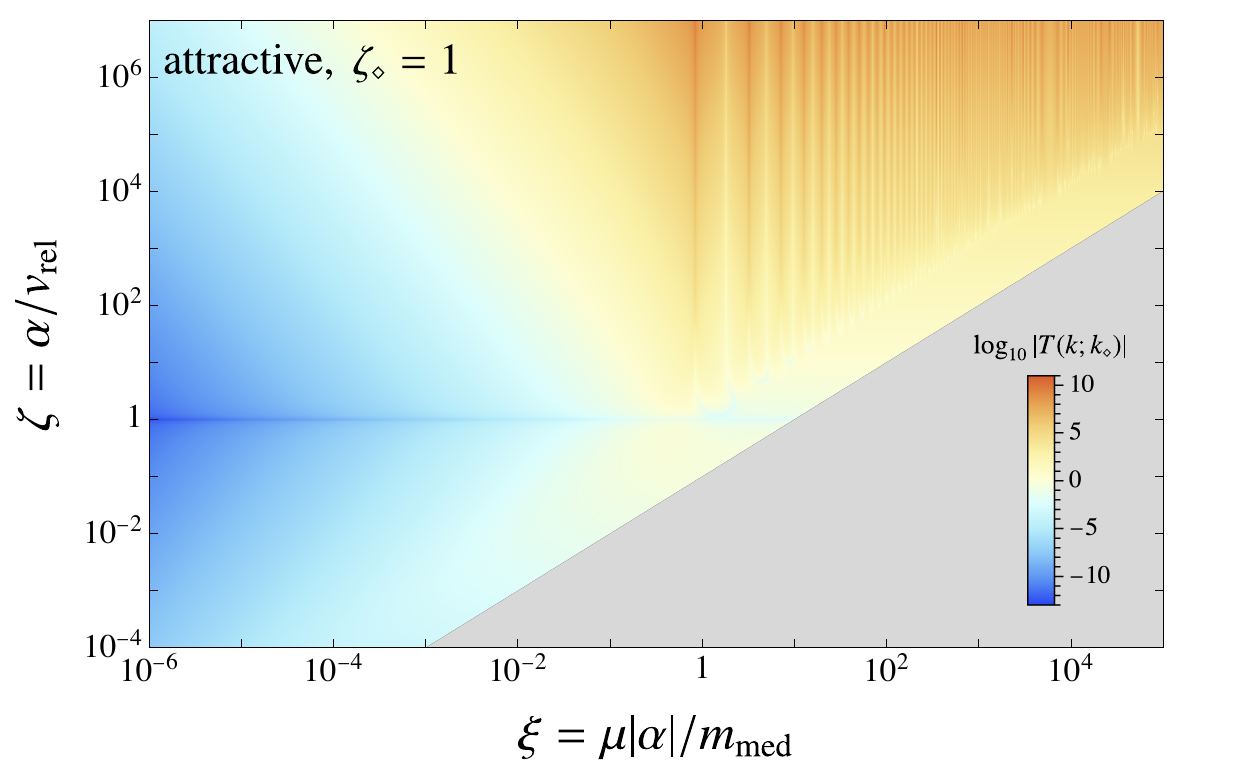}
    \caption{$|T(k;k_\diamond)|$ as a function of $\xi$ and $\zeta = k_\diamond/k$ for $\zeta_\diamond = 1$ and for an attractive potential. The gray region, defined by $10\, |\zeta|<\xi$, is excluded from our numerical evaluation because approaching the Coulomb limit increases the computational cost significantly.
    }
\label{fig:T_Yukawa_vs_k_grid}
\end{figure}

\Cref{fig:T_Yukawa_vs_k} presents computations of $|T(k;k_\diamond)|$ as a function of the ratio $k/k_\diamond$ for $\zeta_\diamond = 1$ and various $\xi$. Both panels show an attractive potential. The solid purple line shows the Coulomb limit $\xi\to\infty$, given by~\cref{eq:Coulomb_T}. For finite $\xi$, $T(k;k_\diamond)$ tracks the Coulomb limit for $\zeta\lesssim\xi$, i.e., $k/k_\diamond \gtrsim 1/\xi$. \Cref{fig:T_Yukawa_vs_k_grid} shows $|T(k;k_\diamond)|$, again for $\zeta_\diamond = 1$, across a broad range of parameters.

\clearpage
\section{Scattering cross sections and numerical convergence} \label{sec:ScatteringCS}

\subsection{Elastic and angularly weighted cross sections}

Having determined the partial-wave phase shifts, we can now compute the scattering cross sections. While the total elastic cross section measures the overall scattering probability, the redistribution of momentum and kinetic energy between particles also depends on the scattering angle. This motivates considering the transfer and viscosity cross sections, which account for the angular dependence of transport and are particularly relevant for SIDM. We define
\begin{subequations}
\label{eq:CrossSections_ETV_def}
\label[pluralequation]{eqs:CrossSections_ETV_def}
\begin{align}
 \sigma_{\elas}
 &\equiv
 \int \dd\Omega\,\dv{\sigma}{\Omega},
\label{eq:ElasticCrossSectionDefinition}
\\
 \sigma_\T
 &\equiv
 \int \dd\Omega\,(1-\cos\vartheta)
 \dv{\sigma}{\Omega},
\label{eq:TransferCrossSectionDefinition}
\\
 \sigma_\V
 &\equiv
 \int \dd\Omega\,\sin^2\vartheta
 \dv{\sigma}{\Omega}.
\label{eq:ViscosityCrossSectionDefinition}
\end{align}
\end{subequations}
Using the partial-wave expansion of the scattering amplitude, \cref{eq:fk_PWanalysis,eq:fk_PW}, these cross sections can be expressed directly in terms of the partial-wave amplitudes $f_{k,\ell}$, and hence in terms of the phase shifts derived above~\cite[Appendix~A]{Petraki:2026krj}
\begin{subequations}
\label{eq:CrossSections_PW_sym}
\label[pluralequation]{eqs:CrossSections_PW_sym}
\begin{alignat}{2}
\sigma_\elas(k)
&= \sum_{\ell=0}^\infty \sigma_\ell^\elas(k),
\qquad
&\sigma_\ell^\elas(k)
&\equiv
4\pi (2\ell+1)\abs{f_{k,\ell}}^2 ,
\label{eq:CrossSection_Elastic_PW_sym}
\\
\sigma_{\T}(k)
&= \sum_{\ell=0}^\infty \sigma_\ell^{\T}(k),
\qquad
&\sigma_\ell^{\T}(k)
&\equiv
4\pi(\ell+1)
\abs{
f_{k,\ell+1}
-
f_{k,\ell}
}^2 ,
\label{eq:CrossSection_Transfer_PW_sym}
\\
\sigma_{\V}(k)
&= \sum_{\ell=0}^\infty \sigma_\ell^{\V}(k),
\qquad
&\sigma_\ell^{\V}(k)
&\equiv
4\pi
\frac{(\ell+1)(\ell+2)}{2\ell+3}
\abs{f_{k,\ell+2}-f_{k,\ell}}^2 .
\label{eq:CrossSection_Viscosity_PW_sym}
\end{alignat}
\end{subequations}

\subsection{Analytic approximations for Yukawa scattering}

We now compute the Yukawa scattering cross sections from the numerical phase shifts and compare them with the Born and WKB approximations developed below.

Unless stated otherwise, for the remainder of this work we consider distinguishable particles, $\sym_\ell=1$ for all $\ell$. In this case the cross sections of \cref{eqs:CrossSections_PW_sym} reduce to
\begin{subequations}\label{eq:Disting_PW_CS}
\begin{align}
\sigma_\ell^\elas &= \frac{4\pi}{k^2}(2\ell+1)\sin^2\theta_\ell,
\\
\sigma_\ell^\T &= \frac{4\pi}{k^2}(\ell+1)\sin^2(\theta_{\ell+1}-\theta_\ell),
\\
\sigma_\ell^\V &= \frac{4\pi}{k^2}\frac{(\ell+1)(\ell+2)}{2\ell+3}\sin^2(\theta_{\ell+2}-\theta_\ell).
\end{align}
\end{subequations}
The compact form of $\sigma_\ell^\T$ assumes that neighboring partial waves are both allowed. For identical particles only every other $\ell$ survives, so the interference terms in Eq.~\eqref{eq:CrossSection_Transfer_PW_sym} vanish and $\sigma_\T=\sigma_\elas$. Equivalently,
$\dd\sigma/\dd\Omega$ is symmetric under $\vartheta\to\pi-\vartheta$.

\subsubsection{Born approximation}

The first Born approximation replaces the exact scattering state in the Lippmann--Schwinger equation with the free incoming state. For a spherically symmetric potential, the resulting scattering amplitude is~\cite{Sakurai:2011zz}
\begin{align}
f_{\vb{k}}^{\Born}(\vartheta)
=
-\frac{2\mu}{q}\int_0^\infty\dd r\ r\ V(r)\ \sin qr,\ \quad q = 2k\sin(\vartheta/2)
.
\end{align}
For the Yukawa potential, \cref{eq:YukawaPotential}, this yields
\begin{align}
\dv{\sigma^{\Born}}{\Omega} = \frac{4\mu^2\alpha^2}{[m_{\med}^2 + 2k^2(1-\cos\vartheta)]^2}
=
\frac{1}{k^2}
\frac{4\zeta^2}{[(\zeta/\xi)^2 + 2(1 - \cos\vartheta)]^2}
.
\end{align}
The elastic and angularly weighted cross sections follow directly,
\begin{subequations}
\begin{align}
\sigma_{\rm elas}^{\Born} 
&=
\frac{4 \pi}{k^2}
\frac{4{\xi}^4}{\qty(\zeta ^2 + 4 {\xi}^2)},
\\[1ex]
\sigma_\T^{\Born} 
&=
\frac{2 \pi}{k^2}
\zeta^2 
\qty[
\ln(
  \frac{4 {\xi}^2}{\zeta^2}+1)
-\frac{4 {\xi}^2}{\zeta^2+4 {\xi}^2}
],
\\[1ex]
\sigma_\V^{\Born} 
&=
\frac{4 \pi}{k^2}\zeta^2 
\qty[
\qty(1 +\frac{\zeta^2}{2{\xi}^2}) 
\ln(\frac{4 {\xi}^2}{\zeta^2}+1)
-2].
\end{align}
\end{subequations}
The Born phase shifts follow from the partial-wave projection,
\begin{align}
\theta_\ell^{\Born}(k)
\equiv \frac{k}{2}\int_{-1}^{1} \dd (\cos\vartheta)\, P_\ell(\cos\vartheta)\ f_{\vb{k}}^{\Born}(\vartheta)
.
\end{align}
For the Yukawa potential, one obtains
\begin{align}\label{eq:Born_Phase_Shift_Yukawa}
\theta_\ell^{{\Born}, Y}(k)
= \zeta\, Q_\ell\qty[1 + \frac{1}{2}\qty(\frac{\zeta}{\xi})^2].
\end{align}
Here $Q_\ell$ is the Legendre function of the second kind. We introduce $\eta>0$ via
\begin{align}\label{eq:eta_def}
\cosh\eta \equiv 1 + \frac{1}{2}\qty(\frac{\zeta}{\xi})^2
\iff
\sinh(\eta/2) = \frac{|\zeta|}{2\xi}.
\end{align}
At sufficiently large $\ell$, \cref{eq:Legendre_Q_Integral_Rep,eq:Q_small_l_approx} yield
\begin{align}\label{eq:Born_Phase_Shift_Large_ell}
\theta_{\ell\to\infty}^{{\Born}, Y}(k)
\approx
\zeta\, 
e^{-(\ell + 1/2)\eta}
\sqrt{\frac{\pi}{2(\ell + 1/2)\sinh\eta}}
.
\end{align}

\subsubsection{WKB approximation}\label{sec:Yukawa_WKB}

In the weak-potential limit, the WKB approximation yields the Yukawa phase shift (see \cref{eq:WKB_Weak_Potential_PS,app:WKB_Approx}),

\begin{align}\label{eq:PS_Y_WKB_Approx}
\theta_\ell^{\WKB, Y}(k)
\approx
\zeta K_0\qty[\qty(\ell + \frac{1}{2})\frac{|\zeta|}{\xi}],
\end{align}
where $K_n(z)$ is the modified Bessel function of the second kind.

In the semiclassical limit, the behavior of the scattering cross sections is governed by the transition between the rapidly oscillating low-$\ell$ regime and the slowly varying large-$\ell$ regime. We therefore introduce characteristic angular momenta $\ell_\star^\WKB$, with $\star=\{\elas,\T,\V\}$, characterizing the transition between these two regimes. The resulting expressions for the scattering cross sections are given below, with details of their derivation provided in \cref{app:WKB_Approx}.

\paragraph{Characteristic angular momenta:} 
The transition scales relevant for the three cross sections are given by 
\begin{subequations}
\begin{align}
\ell_{\elas}^\WKB(\zeta,\xi) 
&\equiv 
\frac{\xi}{|\zeta|} 
K_0^{-1}
\qty(\frac{1}{\sqrt{2}\,\abs{\zeta} }),
\\[1ex]
\ell_{\T}^\WKB(\zeta,\xi)
&\equiv 
\max\qty{
0,
\frac{\xi}{|\zeta|} K_1^{-1}\qty(\frac{\xi}{\sqrt{2}\,\zeta^2})
-
\frac{1}{2}
},\\[1ex]
\ell_{\V}^{\WKB}(\zeta,\xi)
&\equiv
\max\qty{
0,
\frac{\xi}{|\zeta|}
K_1^{-1}\qty(
\frac{\xi}{2\sqrt{2}\,\zeta^2}
)
- 1 
},
\end{align}
\end{subequations}
where $K_n^{-1}(z)$ is the functional inverse of the modified Bessel function of the second kind.

\paragraph{WKB scattering cross sections:}
Using these characteristic angular momenta, the three cross sections can be written in the following form:
\begin{subequations}\label{eq:WKB_CS}
\begin{align}
\frac{\sigma_{\elas}}{\sigma_{\ell = 0}^\U}
&\approx
\frac{1}{2}\, \qty(\ell_{\elas}^\WKB)^2
+ 
\xi^2\ 
K_{\elas}\qty(\frac{|\zeta|\,\ell_{\elas}^\WKB}{\xi}),\label{eq:sigma_elas_WKB}\\
\frac{\sigma_{\T}}{\sigma_{\ell = 0}^\U}
&\approx 
\frac{1}{4}\, \ell_{\T}^\WKB(\ell_{\T}^\WKB + 1)
+
\frac{\zeta^2}{2}\,
K_{\T}\qty[
\qty(\ell_{\T}^{\WKB} + \frac{1}{2})
\frac{
|\zeta|
}{\xi}
],\\[1ex]
\frac{\sigma_{\V}}{\sigma_{\ell = 0}^\U}
&\approx
\frac{1}{8}\ell_{\V}^\WKB(\ell_{\V}^\WKB + 2)
+
\zeta^2\,
K_{\T}\qty[
(\ell_{\V}^{\WKB} + 1)
\frac{
|\zeta|
}{\xi}
],
\end{align}
\end{subequations}
where $K_{\elas}(z), K_{\T}(z)$ are defined in \cref{eq:K_elas_def,eq:K_T_def}. Since \cref{eq:PS_Y_WKB_Approx} is odd in $\zeta$, the cross sections in \cref{eq:WKB_CS} are identical for attractive and repulsive potentials. The difference between the two, present in the full semiclassical treatment~\cite{Colquhoun:2020adl}, enters beyond the weak-potential limit; as \cref{fig:cross_compare} shows, the weak-potential result nevertheless reproduces the scaling of both numerical results, which is sufficient for the comparison here.

\subsection{Numerical results for Yukawa scattering}

\begin{figure}[ht!]
    \centering
    \includegraphics[width=0.475\linewidth]{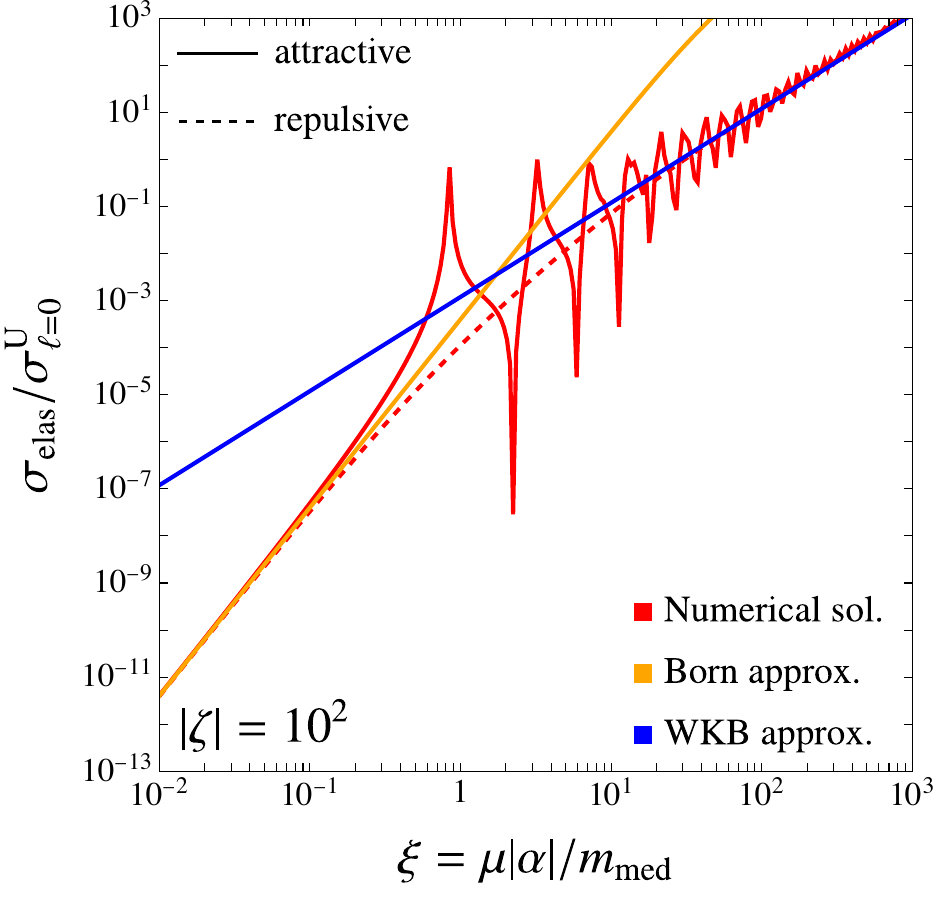}
    \includegraphics[width=0.475\linewidth]{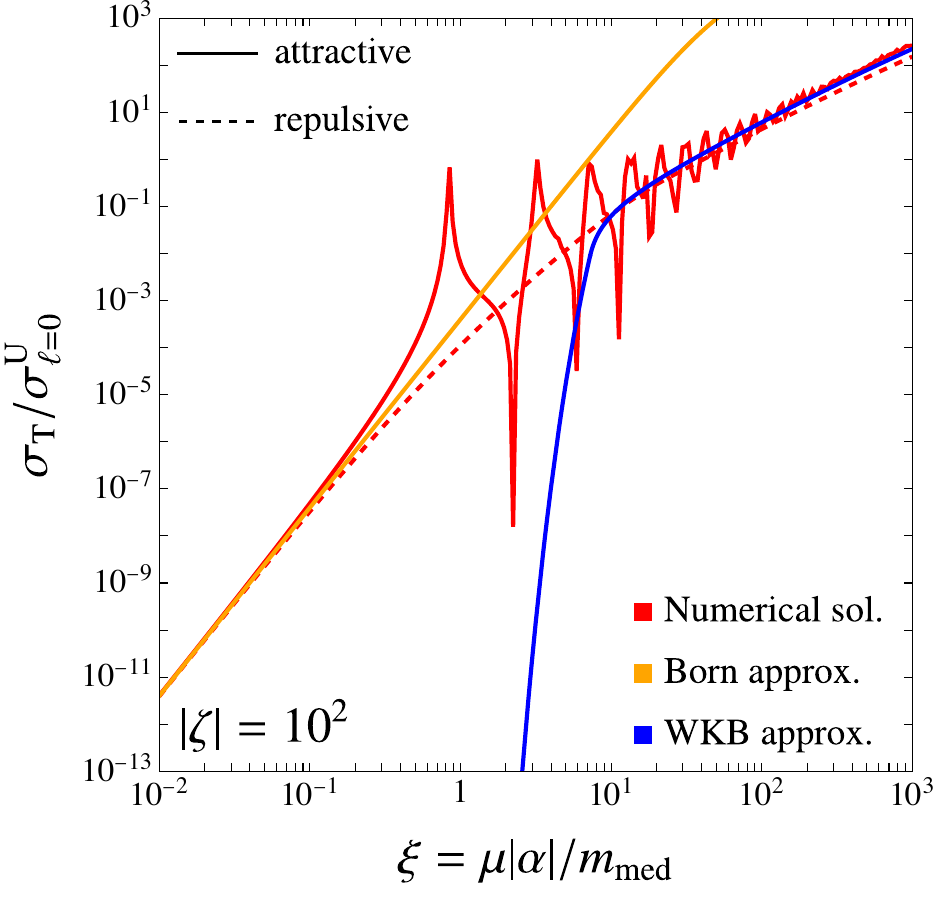}\\[2ex]
    \includegraphics[width=0.475\linewidth]{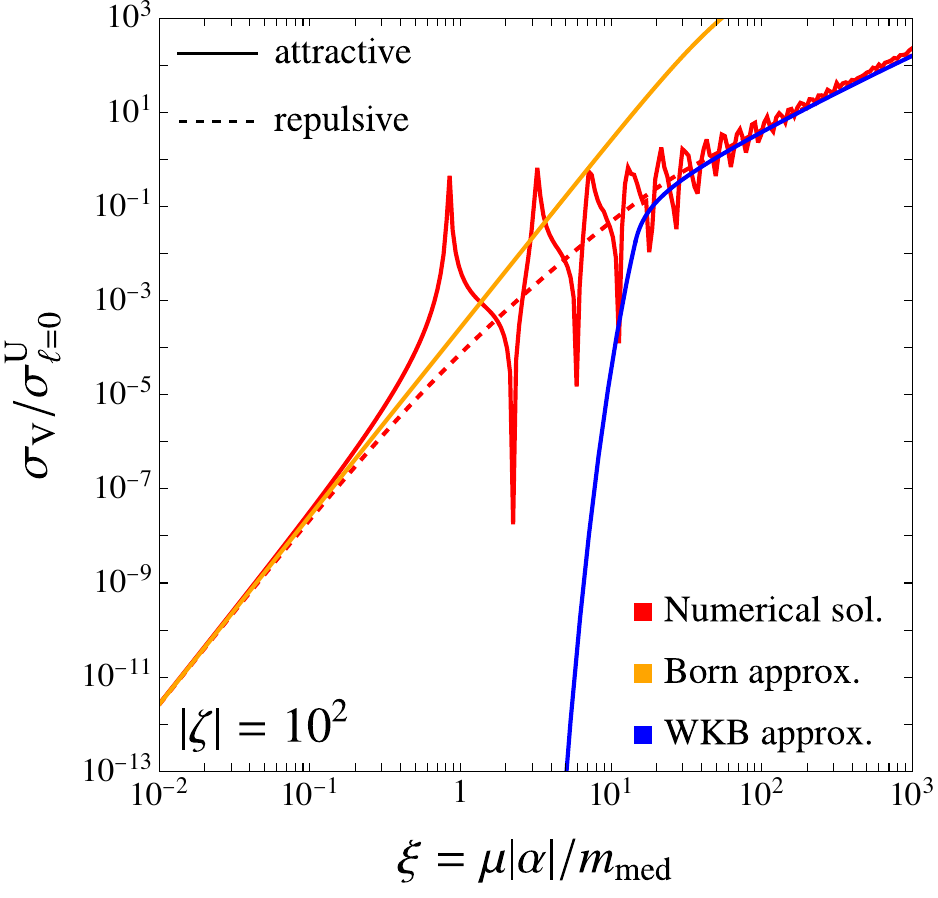}
    \caption{
    The elastic (\textit{upper left}), transfer (\textit{upper right}), and viscosity (\textit{lower}) scattering cross sections as a function of $\xi$. The cross sections include partial waves up to $\ell=100$, $99$, and $98$ for elastic, transfer, and viscosity scattering, respectively. The numerical results are shown together with the Born and WKB approximations presented in the text. The transfer cross section uses the form valid for distinguishable particles ($\sym_\ell=1$); for identical particles $\sigma_{\elas}$ and $\sigma_{\T}$ coincide.
    }
    \label{fig:cross_compare}
\end{figure}

\begin{figure}[ht!]
    \centering
    \includegraphics[width=0.725\linewidth]{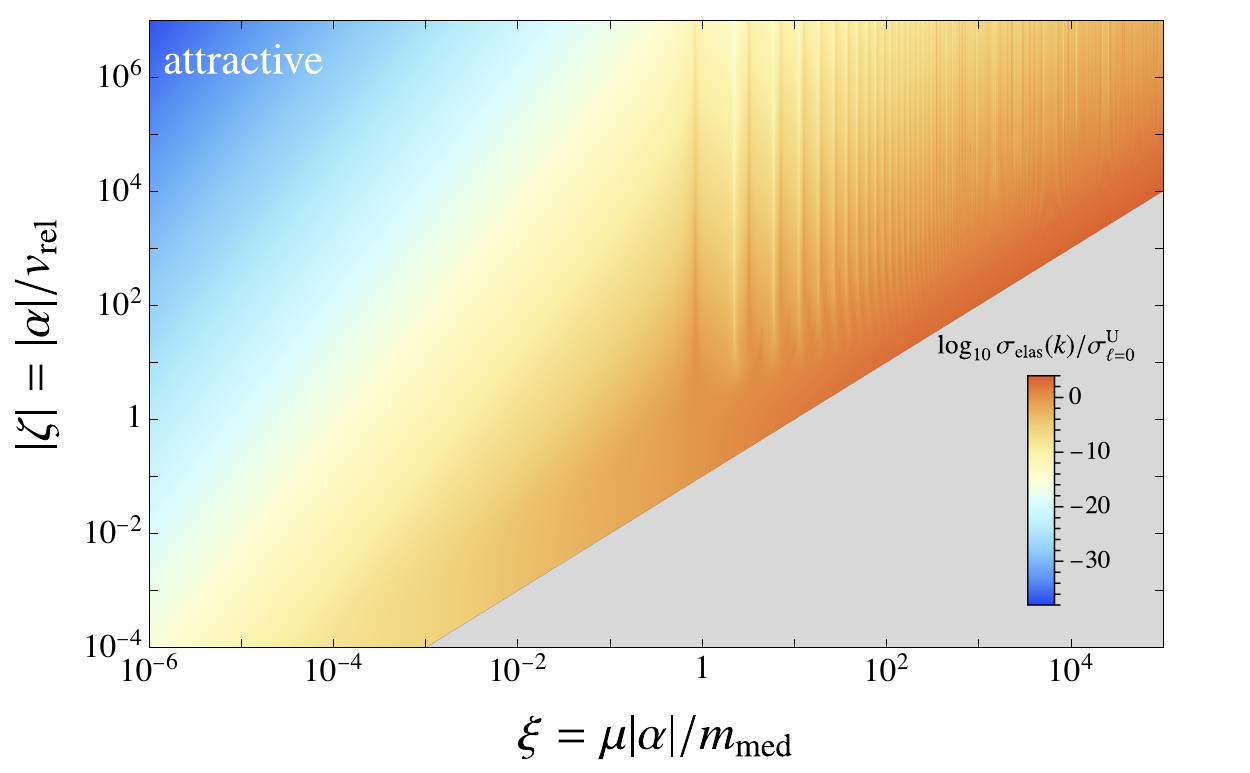}
    \includegraphics[width=0.725\linewidth]{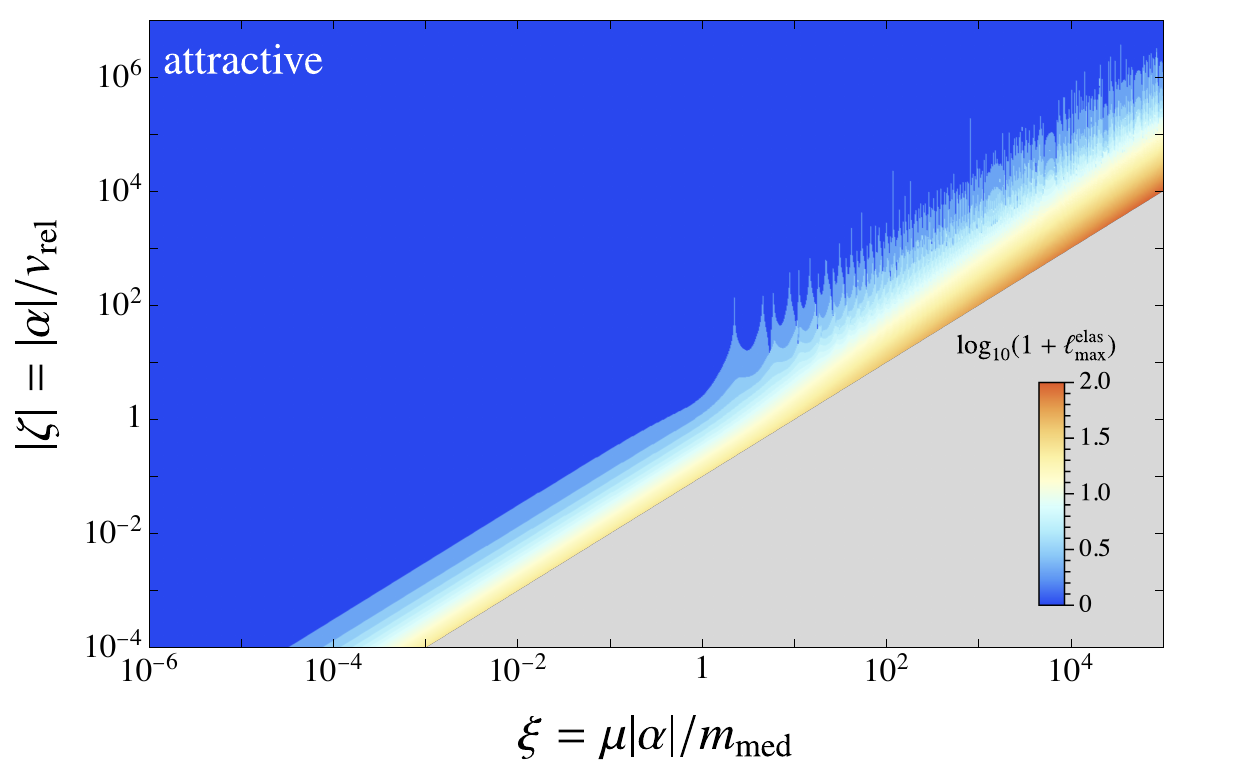}
    \caption{(\textit{upper}) The elastic cross section for an attractive Yukawa potential as a function of $\zeta$ and $\xi$, including partial waves up to $\ell=100$. The gray region, $10\,|\zeta|<\xi$, is excluded as approaching the Coulomb limit requires increasingly many partial waves. (\textit{lower}) Smallest $\ell_{\max}$ beyond which the geometric-tail estimate of the relative truncation error, \cref{eq:Full_Trunc_Error_Est}, is at most $10^{-2}$ for all larger $\ell$ computed; the color bar shows $\log_{10}(1+\ell_{\max}^{\elas})$. We assumed $\sym_\ell=1$ for all $\ell$.}
    \label{fig:Full_Scan_Elas_A}
\end{figure}

\begin{figure}[ht!]
    \centering
    \includegraphics[width=0.725\linewidth]{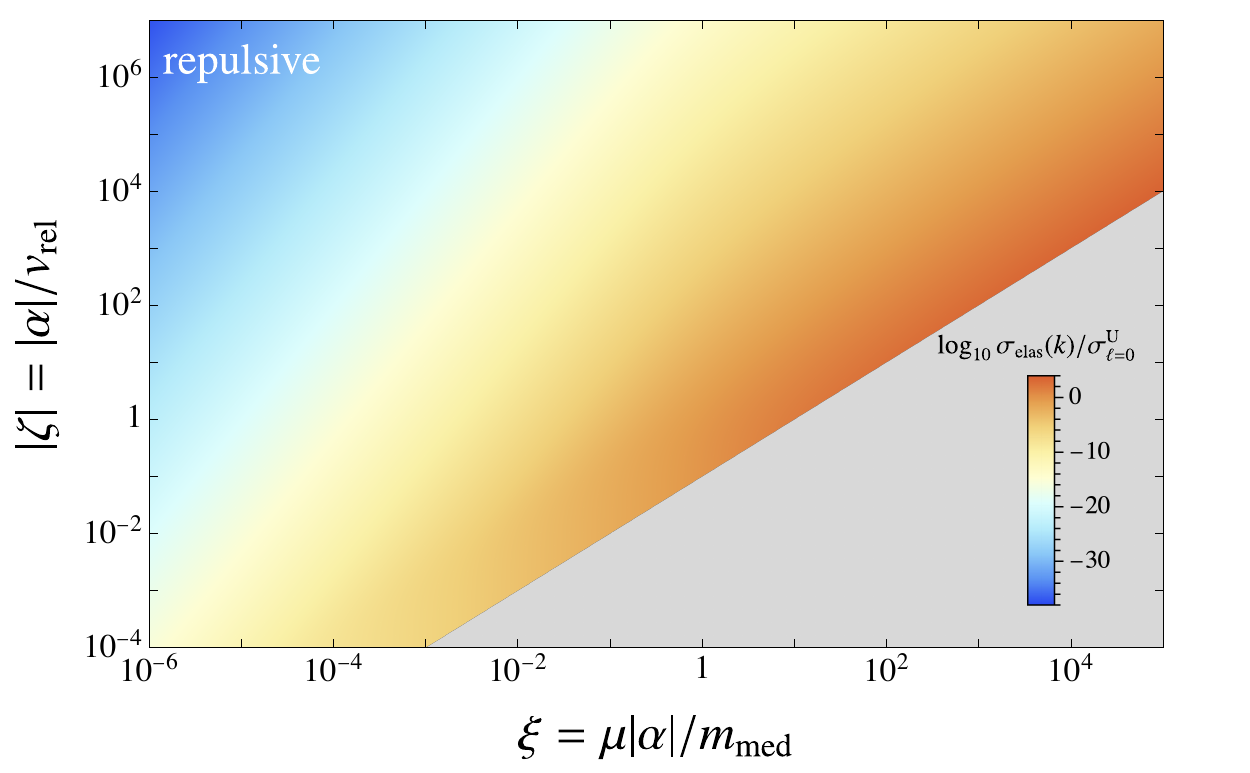}
    \includegraphics[width=0.725\linewidth]{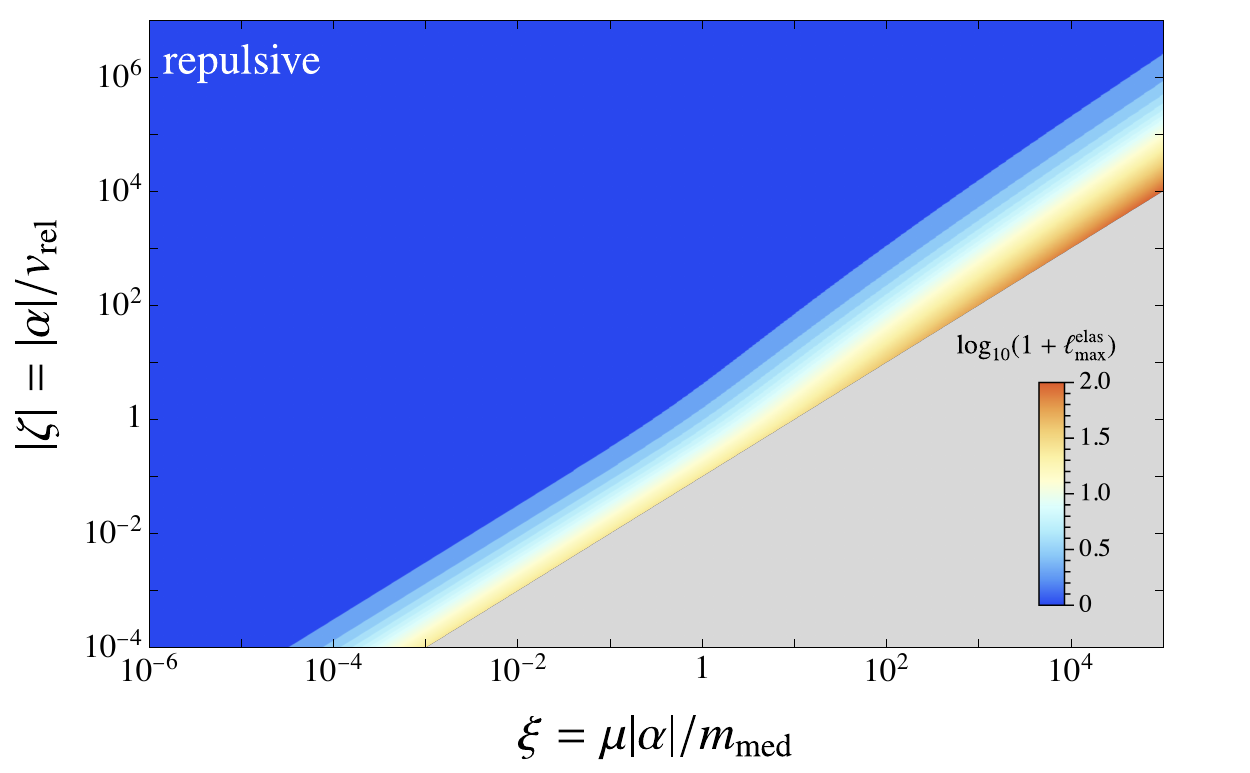}
    \caption{Same as~\cref{fig:Full_Scan_Elas_A}, but for a repulsive Yukawa potential.}
    \label{fig:Full_Scan_Elas_R}
\end{figure}

Examples of the elastic, transfer, and viscosity cross sections are shown in \cref{fig:cross_compare}. We include partial waves up to $\ell=100$, $99$, and $98$ for the elastic, transfer, and viscosity cross sections, respectively. The numerical results are compared with the Born and WKB approximations discussed above, illustrating the regimes in which each approximation provides an accurate description of the numerical scattering calculation. \Cref{fig:Full_Scan_Elas_A,fig:Full_Scan_Elas_R} show the elastic cross section across the $(\zeta,\xi)$ parameter space. We also determine the corresponding $\ell_{\max}$ as the smallest $\ell$ beyond which the geometric-tail estimate, \cref{eq:Full_Trunc_Error_Est}, remains below $10^{-2}$ for all larger $\ell$ computed. Requiring the criterion to hold for all larger $\ell$, rather than at a single $\ell$, removes the spurious dips of the estimator discussed in \cref{sec:PW_Trunc_Conv}. Since the tabulated phase shifts extend to $\ell=100$, the criterion is evaluated up to the last available increment of each partial-wave sum.

\clearpage

\clearpage

\clearpage

\subsection{Partial-wave truncation and convergence}\label{sec:PW_Trunc_Conv}

Here we will examine the truncation error resulting from introducing a cutoff in the partial-wave expansions of~\cref{eqs:CrossSections_PW_sym}. We define
\begin{align}
\sigma_{\star}^{(\ell_{\max})} 
\equiv 
\sum_{\ell = 0}^{\ell_{\max}}
\sigma_{\ell,\star}.
\end{align}
The associated truncation error is 
\begin{align}
\Delta\sigma_{\star}^{(\ell_{\max})} 
\equiv 
\sigma_{\star}
-
\sigma_{\star}^{(\ell_{\max})}
=
\sum_{\ell = \ell_{\max} + 1}^{\infty}
\sigma_{\ell,\star}.
\end{align}
For later use, we will also define the \textit{incremental relative} truncation error
\begin{align}\label{eq:Relative_Trunc_Error}
\varepsilon_{\ell_{\max}}^{\star}
\equiv 
\frac{1}{\sigma_{\star}^{(\ell_{\max})} }
\abs{\sigma_{\star}^{(\ell_{\max} + 1)}  - \sigma_{\star}^{(\ell_{\max})} }
.
\end{align}

\subsubsection*{Elastic cross section} 

Assuming $\ell_{\max}$ is sufficiently large, the Born approximation for the phase shift applies, irrespective of the values of $(\zeta,\xi)$. The truncation error for the elastic cross section becomes
\begin{align}\label{eq:Elas_Truc_Error_Approx}
\frac{\Delta\sigma_{\elas}^{(\ell_{\max})}}{\sigma_{\ell = 0}^\U}
\approx
\sum_{\ell = \ell_{\max} + 1}^{\infty}
(2\ell + 1)\qty[\theta_\ell^{{\Born}, Y}(k)]^2,
\end{align}
where we assumed that $\sin\theta_\ell\sim \theta_\ell$. Inserting \cref{eq:Born_Phase_Shift_Large_ell} into \cref{eq:Elas_Truc_Error_Approx} yields,
\begin{align}\label{eq:Elastic_Trunc_Error}
\frac{\Delta\sigma_{\elas}^{(\ell_{\max})}}{\sigma_{\ell = 0}^\U}
&\approx
\frac{\pi \zeta^2}{\sinh\eta}e^{-\eta}
\sum_{\ell = \ell_{\max} + 1}^{\infty} \qty(e^{-2\eta})^{\ell}\nonumber \\[1ex]
&=
\frac{\pi \zeta^2}{2\sinh^2\eta}\ e^{-2(\ell_{\max} + 1)\eta} =
\frac{2\pi\xi^4}{4\xi^2 + \zeta^2}\ e^{-2(\ell_{\max} + 1)\eta},
\end{align}
where, in the second equality, we recognize the geometric series. In the left panel of~\cref{fig:Error_Est}, we compare the numerically determined truncation error with the estimate given in \cref{eq:Elastic_Trunc_Error}. 

\subsubsection*{Transfer cross section} 

For $\ell \gg (2\eta)^{-1}$,
\begin{align}
\theta_{\ell + 1}^{\Born, Y} - \theta_{\ell}^{\Born, Y} \approx \theta_{\ell}^{\Born, Y}(e^{-\eta} - 1)
.
\end{align}
The truncation error for the transfer cross section is then,
\begin{align}
\frac{\Delta\sigma_{\T}^{(\ell_{\max})}}{\sigma_{\ell = 0}^\U}
\approx
(1 - e^{-\eta})^2
\sum_{\ell = \ell_{\max} + 1}^{\infty}(\ell + 1)\qty[\theta_{\ell}^{\Born, Y}]^2
\approx
\frac{1}{2}
(1 - e^{-\eta})^2\ 
\frac{\Delta\sigma_{\elas}^{(\ell_{\max})}}{\sigma_{\ell = 0}^\U},
\end{align}
where we used the fact that $\ell + 1 \sim (2\ell + 1)/2$ for $\ell\gg 1$. Using \cref{eq:Elastic_Trunc_Error}, one finds
\begin{align}\label{eq:T_Truc_Error_Approx}
\frac{\Delta\sigma_{\T}^{(\ell_{\max})}}{\sigma_{\ell = 0}^\U}
\approx
\frac{\pi \zeta^2}{2\sinh\eta}
\tanh(\eta/2)
e^{-(2\ell_{\max} + 3)\eta}
.
\end{align}

\subsubsection*{Viscosity cross section} 

Similar to the transfer cross section, for $\ell \gg (2\eta)^{-1}$,
\begin{align}
\theta_{\ell + 2}^{\Born, Y} - \theta_{\ell}^{\Born, Y} \approx \theta_{\ell}^{\Born, Y}(e^{-2\eta} - 1)
.
\end{align}
Following a similar procedure as with the transfer cross section,
\begin{align}
\frac{\Delta\sigma_{\V}^{(\ell_{\max})}}{\sigma_{\ell = 0}^\U}
&\approx
(1 - e^{-2\eta})^2
\sum_{\ell = \ell_{\max} + 1}^{\infty}
\frac{(\ell + 1)(\ell + 2)}{(2\ell + 3)}
\qty[\theta_{\ell}^{\Born, Y}]^2 \nonumber \\ 
&\approx
\frac{1}{4}
(1 - e^{-2\eta})^2\ 
\frac{\Delta\sigma_{\elas}^{(\ell_{\max})}}{\sigma_{\ell = 0}^\U},
\end{align}
where we approximated $(\ell + 1)(\ell + 2)/(2\ell + 3)\sim (2\ell + 1)/4$ for $\ell \gg 1$. Combining this result with \cref{eq:Elastic_Trunc_Error}, one finds
\begin{align}
\frac{\Delta\sigma_{\V}^{(\ell_{\max})}}{\sigma_{\ell = 0}^\U}
\approx
\frac{\pi\zeta^2}{2} e^{-2(\ell_{\max} + 2)\eta}
.
\end{align}

\subsubsection*{Summary} 

The above results show that the truncation errors for all three cross sections are exponentially suppressed with $\ell_{\max}$, with the suppression controlled by the parameter $\eta$. In particular, for $|\zeta|/\xi\ll 1$, we have $\eta\simeq |\zeta|/\xi$, so the exponential suppression becomes weak and a parametrically large number of partial waves is required to achieve a given accuracy. Conversely, when $|\zeta|/\xi\gg 1$, $\eta$ becomes large and the partial-wave sum converges rapidly. Among the three cross sections, the elastic cross section generally exhibits the largest truncation error, since its partial-wave sum lacks the additional angular-difference suppression present in the transfer and viscosity cross sections.

For the parameter space shown in~\cref{fig:Full_Scan_Elas_A,fig:Full_Scan_Elas_R} and included in the accompanying dataset, $\eta$ is smallest along the boundary $\xi = 10\,|\zeta|$, where it takes the constant value $\eta = 2\,\mathrm{arcsinh}(1/20)\simeq 0.1$. Convergence in $\ell_{\max}$ is therefore slowest there. Along this boundary, the truncation error in~\cref{eq:Elastic_Trunc_Error} grows as $\zeta^2$ at fixed $\ell_{\max}$, faster than $\sigma_{\elas}/\sigma^{\U}_{\ell=0}\propto(\ln|\zeta|)^2$ [\cref{eq:sigma_elas_WKB}], so the relative error is largest at $(\zeta,\xi)=(10^{4},10^{5})$.

\begin{figure}[t!]
    \centering
    \includegraphics[width=0.49\linewidth]{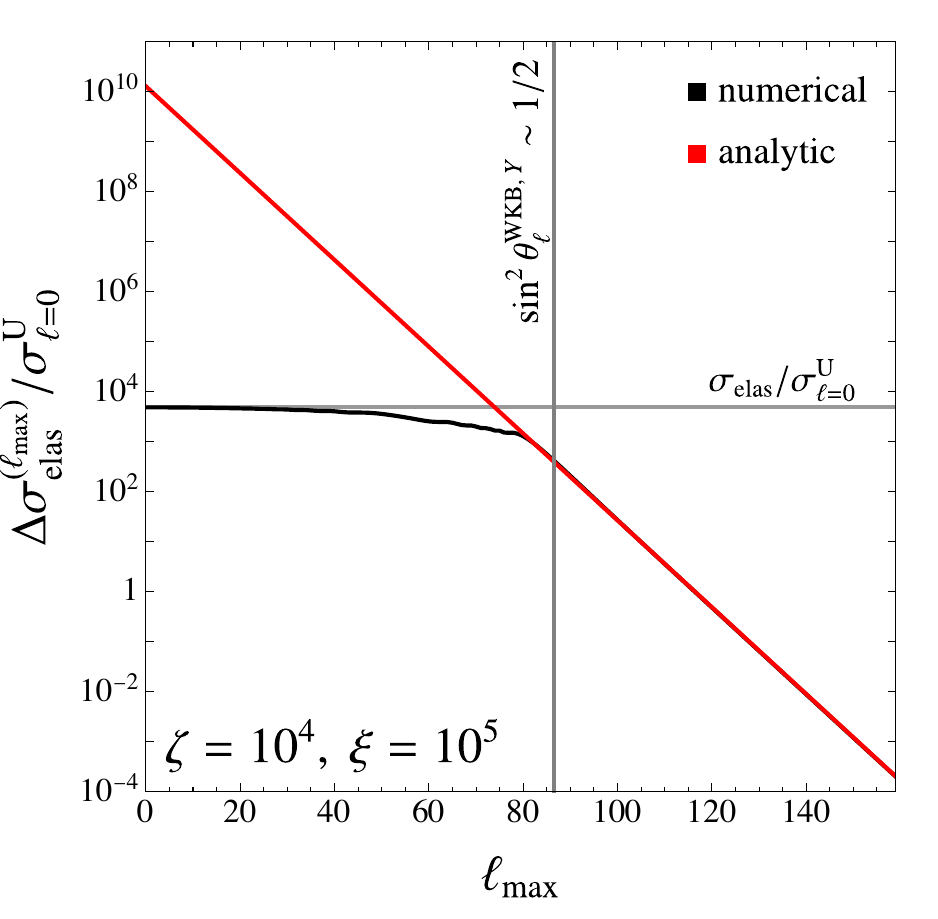}
    \includegraphics[width=0.49\linewidth]{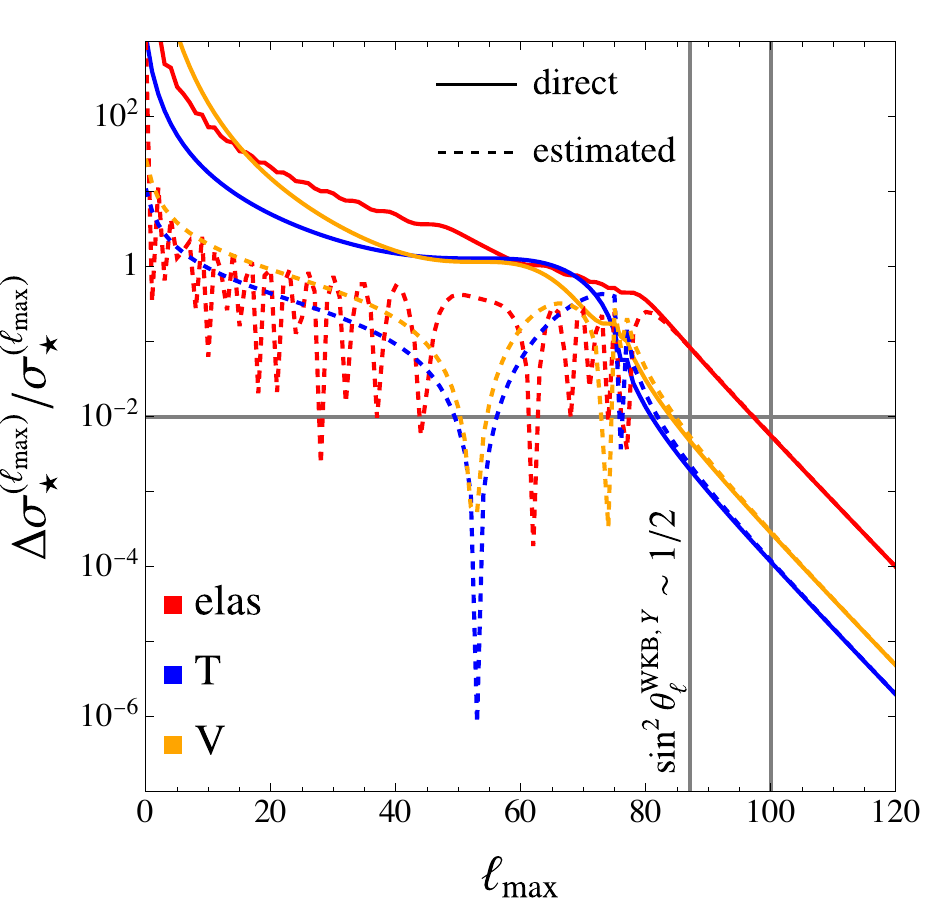}
    \caption{(\textit{left}) A comparison of the elastic cross-section truncation error obtained numerically and analytically using~\cref{eq:Elastic_Trunc_Error}. The horizontal line indicates the asymptotic value of the elastic cross section, while the vertical line marks the point at which the WKB-approximated phase shift becomes of order unity. (\textit{right}) The \textit{full relative} truncation error for the elastic, transfer, and viscosity cross sections, computed directly (solid) and estimated via \cref{eq:Full_Trunc_Error_Est} (dashed). The difference between the cross sections reflects the additional suppression of the transfer and viscosity cross sections arising from the angular differences of neighboring partial waves. Vertical lines mark $\ell_{\elas}^{\WKB}$ and $\ell = 100$; the horizontal line marks the $10^{-2}$ tolerance.
}
    \label{fig:Error_Est}
\end{figure}


Since $\varepsilon^{\star}_{\ell_{\max}}$ compares only the next contribution to the partial sum, it underestimates the \textit{full relative} truncation error. Once the partial-wave contributions decrease geometrically, the two are related by the geometric-tail factor,
\begin{align}\label{eq:Full_Trunc_Error_Est}
\frac{\Delta\sigma_{\star}^{(\ell_{\max})}}{\sigma_{\star}^{(\ell_{\max})}}
\approx
\frac{\varepsilon^{\star}_{\ell_{\max}}}{1 - e^{-2\eta}},
\qquad
\ell_{\max}
\gtrsim 
\ell_{\elas}^{\WKB},
\end{align}
where $(1-e^{-2\eta})^{-1}\simeq 5.5$ for $\eta\simeq 0.1$. For $\ell_{\max}\lesssim\ell_{\elas}^{\WKB}$, the phase shifts are not small and the partial-wave contributions do not decrease monotonically. In this regime, $\varepsilon^{\star}_{\ell_{\max}}$ can be accidentally small whenever a single contribution nearly vanishes, and \cref{eq:Full_Trunc_Error_Est} does not apply.

To test \cref{eq:Full_Trunc_Error_Est}, we compute the full relative truncation error directly at the most demanding grid point, $(\zeta,\xi)=(10^4,10^5)$, using the numerical phase shifts up to $\ell=164$ and the Born phase shifts, \cref{eq:Born_Phase_Shift_Yukawa}, up to $\ell=1000$. The right panel of \cref{fig:Error_Est} shows that the direct result decreases monotonically. The estimate shows spurious dips below $\ell_{\elas}^{\WKB}\simeq87$ and agrees with the direct result within a few percent above it. For the sums up to $\ell=100$, $99$, and $98$, the truncation errors of the elastic, transfer, and viscosity cross sections are $5.6\times10^{-3}$, $1.4\times10^{-4}$, and $4.3\times10^{-4}$, respectively, all below~$1\%$.

\clearpage

\section{Conclusions} \label{sec:Conclusions}

We have presented a formulation of the variable-phase method for nonrelativistic scattering that treats the regular and irregular solutions of the radial Schr\"odinger equation within a single set of first-order equations. The asymptotic coefficient functions of the regular solution determine both the phase shifts and the Sommerfeld factors, according to~\cref{eq:PhaseShift_Tan_Def,eq:VPS_Sommerfeld}. The irregular solution, obtained by integrating the same equations inward, allows us to extract the finite part of the projected $s$-wave Green's function, $T(k;k_\diamond)$, using~\cref{eq:T_VPA}; this quantity enters in the unitarization of contact-like inelastic processes. Since the singular behavior of the potential at the origin is known, the divergent contribution can be subtracted analytically, and $T(k;k_\diamond)$ follows from the linear coefficient of the irregular solution evaluated separately at each momentum. This procedure makes $T(k;k_\diamond)$ available for potentials without closed-form solutions.

Applied to the Yukawa potential, the method addresses two important challenges in numerical calculations: the choice of integration boundaries and the convergence of the partial-wave expansion. The characteristic length scales span many orders of magnitude across parameter space, rendering fixed integration boundaries unreliable in certain regimes. We therefore determine the integration domain from the physical scales of the problem, as prescribed by~\cref{eq:xmin_numerical,eq:xmax_numerical}. We also derive analytic estimates of the partial-wave truncation error, showing that convergence slows as the parameter $\eta = 2\, {\rm arcsinh}[|\zeta|/(2\xi)]$ decreases, particularly in the semiclassical regime $|\zeta| \ll \xi$. Throughout the tabulated parameter space, the partial-wave sums up to $\ell=100$, $99$, and $98$ keep the truncation error below $1\%$ for the elastic, transfer, and viscosity cross sections, respectively.

The accompanying code and tabulated data provide phase shifts and Sommerfeld factors for $\ell\in[0,100]$, the total elastic, transfer, and viscosity cross sections, and $T(k;k_\diamond)$. The tabulated parameter space spans the Born and resonant regimes and extends to the onset of the semiclassical and Coulomb-like regimes. These data have already been used in companion studies of self-interacting dark matter that investigate how inelastic processes, specifically contact-like annihilation~\cite{Petraki:2026krj} and radiative bound-state formation~\cite{Flores:2026jsf}, modify elastic scattering through the feedback required by unitarity. 

The variable-phase formulation extends to other real, local, central potentials satisfying the condition in~\cref{eq:Potential_ShortRangeAssumption}, with the extraction of $T(k;k_\diamond)$ additionally requiring the short-distance behavior specified in~\cref{eq:Potential_LaurentExpansion}. The framework thus enables reliable nonperturbative calculations of long-range scattering across a broad parameter space relevant for dark-matter phenomenology, while providing the inputs needed for the self-consistent unitarization of contact-like inelastic processes.

\newpage

\section*{Acknowledgments}

This work was supported by the European Union’s Horizon 2020 research and innovation program under grant agreement No 101002846, ERC CoG CosmoChart. MMF is supported through a FRIPRO grant of the Norwegian Research Council (project ID 353561 ‘DarkTurns’).

\section*{Data and code availability}
The \texttt{Mathematica} code is openly available at \url{https://github.com/mmfloresnm/YukawaVPS}, and the tabulated data are archived on Zenodo at \url{https://doi.org/10.5281/zenodo.23475703}~\cite{Flores:2026data}. The repository includes a user guide documenting the parameter grid, data format, and interpolation functions.

\clearpage

\begin{appendices}

\section{Conventions and formulas}\label{app:Formulas}

\subsubsection*{Orthogonality relations}

We use the standard orthogonality relation for Legendre polynomials,
\begin{align}
\label{eq:Legendre_Orthogonality}
\int_{-1}^{1} P_{\ell}(z) P_{\ell'}(z)\ \dd z = \frac{2}{2\ell + 1}\delta_{\ell\ell'}
.
\end{align}

\subsubsection*{Legendre function of the second kind}

When evaluating the Legendre function of the second kind, we use the integral representation~\cite[14.12.14]{NIST:DLMF}
\begin{align}\label{eq:Legendre_Q_Integral_Rep}
Q_\ell(\cosh\eta) = \int_0^\infty \frac{\dd t}{(\cosh\eta + \sinh\eta\cosh t)^{\ell + 1}},
\end{align}
which is manifestly real for $\ell \geq 0$ and $\eta > 0$. Of particular interest is the $\ell\to\infty$ limit. For large $\ell$, the integrand is peaked at $t = 0$, where $\cosh\eta + \sinh\eta = e^\eta$. Expanding in $t$, the integral becomes
\begin{align}
Q_\ell(\cosh\eta) 
\sim
e^{-(\ell + 1)\eta}\int_0^\infty\ \dd t\ 
\exp\qty[-\frac{(\ell + 1)\sinh\eta}{2e^{\eta}}\ t^2],
\qquad \ell \gg \max\left(1, \eta^{-1}\right)
.
\end{align}
Evaluating the integral above yields
\begin{align}\label{eq:Q_small_l_approx}
Q_\ell(\cosh\eta) \sim e^{-(\ell + 1/2)\eta}
\sqrt{\frac{\pi}{2(\ell + 1/2)\sinh\eta}},
\qquad \ell \gg \max\left(1, \eta^{-1}\right),
\end{align}
where we replaced $(\ell+1)$ with $(\ell+ 1/2)$ in the denominator. The difference induces a correction of $\order{1/\ell}$, which is subleading in the large-$\ell$ approximation.

\subsubsection*{Modified Bessel functions and their functional inverses}

The WKB approximation makes extensive use of the modified Bessel functions of the second kind, $K_n(z)$, and their functional inverses. Of particular utility are the $n = 0, 1$ functions, which have the following limiting behavior:
\begin{subequations}
\begin{align}
K_0(z) 
&\underset{z\to 0}{\simeq}
-\ln\left(\frac{z}{2}\right)-\gammaE,
&\qquad
K_1(z) &\underset{z\to 0}{\simeq} \frac{1}{z},
\\[1ex]
K_0(z) &\underset{z\to \infty}{\simeq}
\sqrt{\frac{\pi}{2z}}\,e^{-z},
&\qquad
K_1(z) &\underset{z\to \infty}{\simeq}
\sqrt{\frac{\pi}{2z}}\,e^{-z},
\end{align}
\end{subequations}
where $\gammaE$ is the Euler--Mascheroni constant. From these limits, we obtain approximate expressions for the functional inverses:
\begin{align}
K_0^{-1}(z) 
\simeq 
\begin{cases}
\frac{1}{2}{\cal W}_0(\pi/z^2), & z \ll 1,\\
2e^{-(\gammaE + z)}, & z \gg 1,
\end{cases}
\qquad
K_1^{-1}(z) 
\simeq 
\begin{cases}
\frac{1}{2}{\cal W}_0(\pi/z^2), & z \ll 1,\\
\frac{1}{z}, & z \gg 1,
\end{cases}
\end{align}
where ${\cal W}_0$ is the Lambert ${\cal W}$ function.

\clearpage

\section{The Coulomb limit}
\label{App:Coulomb}

For comparison with the numerical Yukawa results, it is useful to recall the corresponding analytic quantities for the strict Coulomb problem, $V^{C}(r)=-\alpha/r$. Unlike the Yukawa potential at finite $m_{\med}$, the Coulomb interaction is genuinely long-ranged, and its scattering states do not approach free spherical waves at large distances. Instead, the regular Coulomb solution behaves asymptotically as
\begin{align}
{\cal F}^{C}_{k,\ell}(r)
\propto
\sin\qty[
kr-\frac{\ell\pi}{2}
+\zeta\ln(2kr)
+\theta_\ell^{C}(k)
],
\end{align}
where $\theta_\ell^{C}$ is the constant Coulomb phase shift. The Coulomb phase shift, Sommerfeld factor, and $T(k;k_\diamond)$ are~\cite{Blum:2016nrz,Flores:2025qit},
\begin{subequations}
\label{eq:CoulombResults}
\label[pluralequation]{eqs:CoulombResults}
\begin{align}
\theta_\ell^{C}(k)
&=
\arg\Gamma(1+\ell-\im\zeta)
=
\dfrac{1}{2\im}
\ln\qty[
\dfrac{\Gamma(1+\ell-\im\zeta)}
      {\Gamma(1+\ell+\im\zeta)}
],
\label{eq:Coulomb_PS}
\\[1ex]
S_{\ell}^{C}(k)
&=
\dfrac{2\pi\zeta}{1-e^{-2\pi\zeta}}
\prod_{j=1}^\ell
\left(1+\frac{\zeta^2}{j^2}\right),
\label{eq:Coulomb_Sommerfeld}
\\[1ex]
T^C(k;k_\diamond)
&=
2\zeta
\qty[
\ln\qty(\frac{\zeta}{\zeta_\diamond})
+\Re\qty[H_{+\im\zeta_\diamond}]
-\Re\qty[H_{+\im\zeta}]
].
\label{eq:Coulomb_T}
\end{align}
\end{subequations}
Here $H_z$ denotes the harmonic number analytically continued to the complex plane. For each partial wave, the Coulomb phase shift gives
\begin{align}
(\sigma_\ell^{\elas})^{C}
=
\sym_\ell
\dfrac{4\pi}{k^2}(2\ell+1)
\sin^2\theta_\ell^{C}.
\label{eq:Coulomb_Elastic}
\end{align}
The sum of \cref{eq:Coulomb_Elastic} over partial waves diverges for the strict Coulomb potential, reflecting the forward singularity of Rutherford scattering.

Care is needed when comparing $\theta_\ell^{C}$ with the Yukawa phase shift $\theta_\ell(k)$ defined above through free-wave asymptotics. For any finite $m_{\med}$, the Yukawa interaction becomes exponentially suppressed at sufficiently large distances, and the scattering solution eventually approaches a free spherical wave. As $m_{\med}$ decreases, the Coulomb-like region extends to increasingly large distances before this free-wave regime is reached, causing the phase accumulated relative to a free wave to grow logarithmically. This can be seen directly from the variable-phase equation. For an intermediate point $x_0$ sufficiently large for the Coulomb asymptotic form to apply, while still satisfying $x_0\ll k/m_{\med}$, the solution has accumulated the Coulomb phase $\theta_\ell^{C}+\zeta\ln(2x_0)$ by $x_0$, while at larger $x$ the rapidly oscillating factor in the variable-phase equation may be averaged. Thus,
\begin{align}
\theta_\ell(k)
&\simeq
\theta_\ell^{C}(k)
+\zeta\ln(2x_0)
+\zeta\int_{x_0}^{\infty}
\frac{\dd x}{x}\,
e^{-(m_{\med}/k)x}
\nonumber\\
&=
\theta_\ell^{C}(k)
+\zeta\ln(2x_0)
+\zeta E_1\qty(\frac{m_{\med}x_0}{k})
\nonumber\\
&=
\theta_\ell^{C}(k)
+\zeta
\qty[
\ln\qty(\frac{2k}{m_{\med}})
-\gamma_E
]
+o(1),
\qquad
\frac{m_{\med}}{k}\to0,
\label{eq:Yukawa_Coulomb_PhaseRelation}
\end{align}
where $E_1(z)$ is the exponential integral and we used
$E_1(z)=-\gamma_E-\ln z+o(1)$ for $z\to0$. The limit in \cref{eq:Yukawa_Coulomb_PhaseRelation} is understood at fixed $\ell$ and $\zeta$, on a continuous branch of the phase shift, and agrees with the standard screened-Coulomb result for exponential screening~\cite{Taylor:1973da,Yamaguchi:2003sz}. The logarithmically divergent contribution is independent of $\ell$. Consequently, the Yukawa phase shift does not converge pointwise to $\theta_\ell^{C}$ as $\xi\to\infty$, and individual Yukawa partial-wave elastic cross sections develop increasingly rapid oscillations rather than approaching \cref{eq:Coulomb_Elastic} pointwise. The common divergent contribution cancels, however, in differences of partial-wave phase shifts. By contrast, the Sommerfeld factor and $T(k;k_\diamond)$ possess the well-defined Coulomb limits given above.

\clearpage

\section{The WKB approximation}\label{app:WKB_Approx}

In this appendix, we briefly examine the WKB approximation for the Yukawa potential. For a more comprehensive treatment, see Ref.~\cite{Colquhoun:2020adl}. In the semiclassical regime, where the de Broglie wavelength is much smaller than the characteristic scale over which the potential varies, the reduced radial wave function is well approximated by the WKB solution~\cite{Landau:1991wop}
\begin{align}\label{eq:WKB_Def}
u_{k,\ell}(x) 
\simeq
u_{k,\ell}^\WKB(x)
\propto
\frac{1}{{\cal K}_\ell^{1/4}(x)}
\sin(
\int_{x_{\cl}}^{x}{\cal K}_\ell^{1/2}(x')\ \dd x' + \frac{\pi}{4} 
),
\end{align}
where 
\begin{align}
{\cal K}_\ell(x) \equiv 1 -
\frac{(\ell + 1/2)^2}{x^2} - \tilde{V}(x)
=
\bar{\cal K}_\ell(x)
- \tilde{V}(x)
,
\end{align}
and $x_{\rm cl}$ is the classical turning point defined via ${\cal K}_\ell(x_{\rm cl}) = 0$. We have also defined $\bar{\cal K}_\ell$, corresponding to the case where $\tilde{V}(x) = 0$. Taking the far-field limit of~\cite[Eq. 127.4]{Landau:1991wop},

\begin{align}\label{eq:WKB_PhaseShift_Def}
\theta_\ell^\WKB(k)
\equiv
\frac{\pi}{2}\qty(\ell + \frac{1}{2})
+
\qty[
\int_{\xcl}^{\infty}
\qty[{\cal K}_\ell^{1/2}(x')
-1
]\ \dd x'
-
\xcl
]
.
\end{align}

\subsubsection*{Weak-potential limit}

\Cref{eq:WKB_PhaseShift_Def} provides a convenient means of rapidly computing the phase shift in the semiclassical regime. In general, however, its evaluation still requires numerical integration. In order to obtain analytic results, particularly for the Yukawa potential, we consider the weak-potential limit. To do so, we expand the first term of the integrand in \cref{eq:WKB_PhaseShift_Def} to give
\begin{align}\label{eq:WKB_Func_Expand}
{\cal K}_\ell^{1/2}(x)
\simeq
\bar{\cal K}_\ell^{1/2}(x)
- 
\frac{\tilde{V}(x)}{2\bar{\cal K}_\ell^{1/2}(x)} + {\cal O}(\tilde{V}^2)
.
\end{align}
At the same time, we can expand the classical turning point $\xcl = \bxcl + \delta \xcl$ where $\bar{\cal K}_\ell( \bxcl ) = 0$. Specifically, $\bxcl = \ell + 1/2$. Requiring ${\cal K}_\ell(\xcl) = 0$ and expanding to linear order in $\tilde{V}(x)$ and $\delta \xcl$, one obtains
\begin{align}\label{eq:ClassicalTP_Expand}
\delta \xcl
=
\frac{\tilde{V}(\bxcl)}{\bar{\cal K}_\ell^{\prime}(\bxcl)}
=
\frac{(\ell + 1/2)}{2}
\tilde{V}(\bxcl)
.
\end{align}
Upon inserting \cref{eq:WKB_Func_Expand,eq:ClassicalTP_Expand} in \cref{eq:WKB_PhaseShift_Def}, the shift $\delta\xcl$ of the lower limit cancels against the explicit $-\xcl$, since ${\cal K}_\ell^{1/2}(\xcl)=0$. The non-uniformity of~\cref{eq:WKB_Func_Expand} near $\bxcl$ yields an integrable singularity and affects only higher orders. We obtain the weak-potential phase shift,
\begin{align}\label{eq:WKB_Weak_Potential_PS}
\theta_\ell^{\rm WP}(k)
\equiv
\theta_\ell^\WKB(k)
\approx
-\frac{1}{2}
\int_{\ell + 1/2}^{\infty}
\frac{x'\ \tilde{V}(x')}{\sqrt{x'^2 - (\ell + 1/2)^2}}\ \dd x',
\end{align}
where, again, we have only retained leading-order terms. For the Yukawa potential, \cref{eq:YukawaPotential_NonDim}, one easily obtains~\cref{eq:PS_Y_WKB_Approx}.

\subsubsection*{Approximating the scattering cross sections}

\begin{figure}[ht!]
    \centering
    \includegraphics[width=0.55\linewidth]{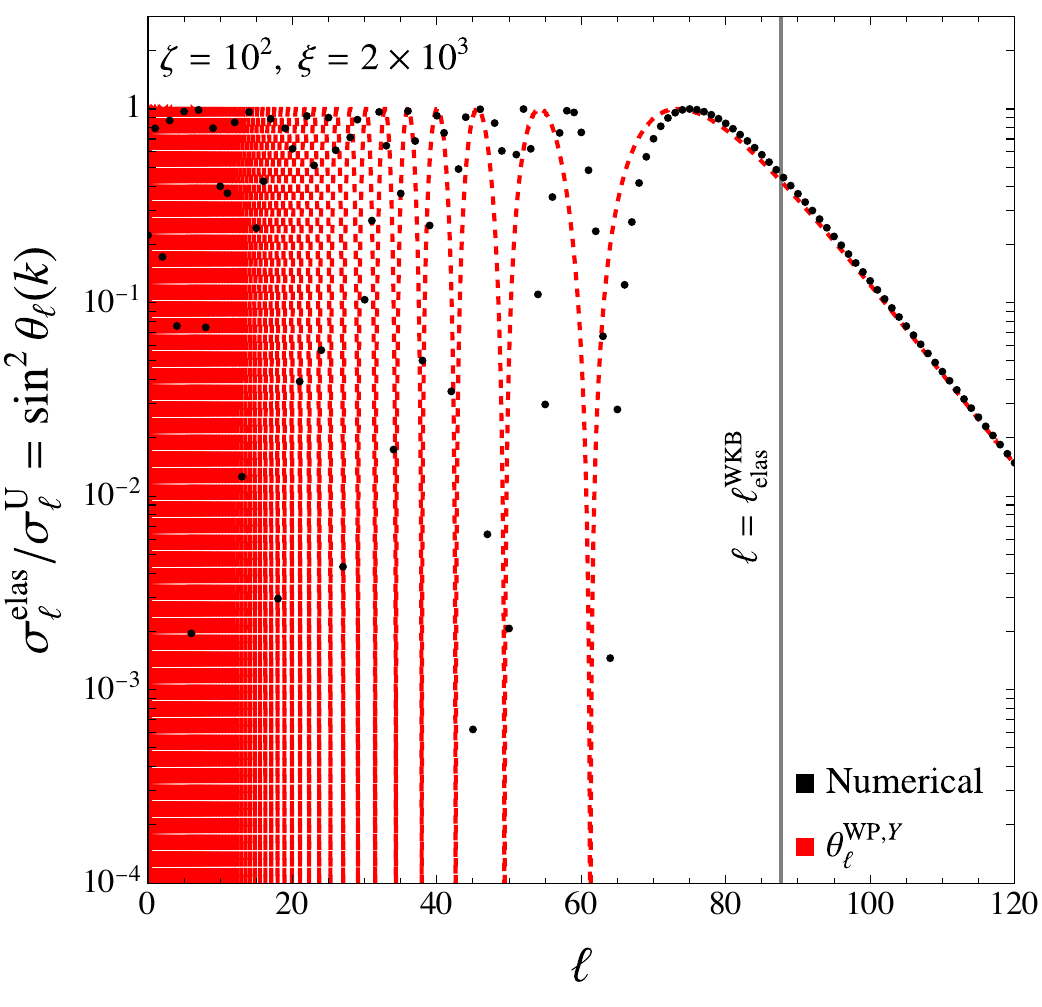}
    \caption{The normalized elastic cross section for $\zeta = 10^2$ and $\xi = 2\times 10^3$ as a function of $\ell$. At small $\ell$, the rapid variation of $\theta_\ell(k)$ produces pronounced oscillations in $\sin^2\theta_\ell(k)$, which are gradually suppressed as $\ell$ increases and the phase shift approaches zero. This behavior motivates the separation of the low- and high-$\ell$ regimes used in the evaluation of~\cref{eq:CrossSections_ETV_WKB}.}
    \label{fig:WKB_PS_Compare}
\end{figure}

In order to obtain analytic expressions for the scattering cross sections, we approximate~\cref{eq:Disting_PW_CS} using the midpoint rule as discussed in~\cite{Colquhoun:2020adl}. One finds,
\begin{subequations}\label{eq:CrossSections_ETV_WKB}
\begin{align}
\frac{\sigma_{\elas}}{\sigma_{\ell = 0}^\U}
&\approx
2\int_{0}^\infty \dd \ell\ 
\ell\ 
\sin^2\theta_{\ell-1/2}(k), 
\label{eq:Elas_CS_WKB}
\\[1ex]
\frac{\sigma_{\T}}{\sigma_{\ell = 0}^\U}
&\approx
\int_{0}^\infty 
\dd \ell\ 
\qty(\ell + \frac{1}{2})\ 
\sin^2\dot{\theta}_{\ell}(k),
\label{eq:T_CS_WKB}
\\[1ex]
\frac{\sigma_{\V}}{\sigma_{\ell = 0}^\U}
&\approx
\frac{1}{2}
\int_{0}^\infty 
\dd \ell\ 
\qty(\ell + 1)\ 
\sin^2[2\dot{\theta}_{\ell + 1/2}(k)],
\label{eq:V_CS_WKB}
\end{align}
\end{subequations}
where $\dot{\theta}_\ell\equiv\dd\theta_\ell/\dd\ell$. To proceed, we note that the phase shift vanishes as $\ell\to\infty$, while at small $\ell$ it varies rapidly, resulting in rapid oscillations of the sine functions appearing in the above integrals. We therefore introduce a characteristic angular momentum $\ell_{\star}^{\WKB}$, with $\star=\{\elas,\T,\V\}$, which separates the rapidly varying low-$\ell$ regime from the slowly varying large-$\ell$ regime. \Cref{fig:WKB_PS_Compare} illustrates this behavior for the elastic cross section.

\paragraph{Elastic cross section:} We employ the integral approximation~\cref{eq:Elas_CS_WKB} and characterize the transition between the rapidly oscillating and slowly varying regimes by a characteristic angular momentum $\ell_{\elas}^{\rm WKB}$, defined by the condition
\begin{align}
\qty[\theta_{\ell_{\elas}^\WKB -1/2}^{\rm WP}(k)]^2
\equiv 
\frac{1}{2}
.
\end{align}
This choice matches the two approximations used below, $\sin^2\theta_{\ell-1/2}\to1/2$ for rapidly varying phase shifts and $\sin^2\theta_{\ell-1/2}\to\theta_{\ell-1/2}^2$ for small ones, so that the approximated integrand is continuous at $\ell=\ell_{\elas}^{\WKB}$(cf.\ \cref{fig:WKB_PS_Compare}). Using \cref{eq:PS_Y_WKB_Approx}, one finds
\begin{align}
\ell_{\elas}^\WKB 
\equiv 
\frac{\xi}{|\zeta|} K_0^{-1}\qty(\frac{1}{\sqrt{2}\,\abs{\zeta} }),
\end{align}
where $K_n^{-1}(z)$ is the inverse of the modified Bessel function of the second kind. The evaluation of this function is discussed in~\cref{app:Formulas}. Starting with~\cref{eq:Elas_CS_WKB}, we split the domain of integration at $\ell=\ell_{\elas}^\WKB$,
\begin{align}\label{eqs:Elas_CS_Approx}
\frac{\sigma_{\elas}}{\sigma_{\ell = 0}^\U}
&\approx
2
\int_{0}^{\ell_{\elas}^\WKB}  \dd \ell\ 
\ell\ 
\sin^2\theta_{\ell-1/2}(k)
+
2
\int_{\ell_{\elas}^\WKB}^\infty \dd \ell\ 
\ell\ 
\sin^2\theta_{\ell-1/2}(k)
\nonumber
\\[1ex]
&\approx
\frac{1}{2}\, 
\qty(\ell_{\elas}^\WKB)^2
+
2
\int_{\ell_{\elas}^\WKB}^\infty 
\dd \ell\ 
\ell\ 
\qty[\theta_{\ell-1/2}(k)]^2,
\end{align}
where we replaced $\sin^2\theta_{\ell-1/2}\to 1/2$ in the first integral, consistent with the definition of $\ell_{\elas}^\WKB$, and $\sin^2\theta_{\ell-1/2}\to\theta_{\ell-1/2}^2$ in the second. Using \cref{eq:PS_Y_WKB_Approx,eqs:Elas_CS_Approx} we obtain
\begin{align}
\frac{\sigma_{\elas}}{\sigma_{\ell = 0}^\U}
\approx
\frac{1}{2}\, \qty(\ell_{\elas}^\WKB)^2
+ 
\xi^2\ 
K_{\elas}\qty(\frac{|\zeta|\,\ell_{\elas}^\WKB}{\xi}),
\end{align}
where
\begin{align}\label{eq:K_elas_def}
K_{\elas}(z)
\equiv
z^2 \qty([K_1(z)]^2-[K_0(z)]^2)
.
\end{align}

\paragraph{Transfer cross section:} In a similar fashion to the elastic cross section, we define
\begin{align}
\qty[\dot{\theta}_{\ell_{\T}^\WKB}^{\rm WP}(k)]^2 \equiv \frac{1}{2}
.
\end{align}
Noting that
\begin{align}
\dot{\theta}_\ell^{\rm WP} = -|\zeta|\frac{\zeta}{\xi}
K_1\qty[\qty(\ell + \frac{1}{2})\frac{|\zeta|}{\xi}],
\end{align}
we solve for $\ell_{\T}^{\WKB}(\zeta,\xi)$. This yields
\begin{align}
\ell_{\T}^{\WKB}(\zeta,\xi)
=
\max\qty{
0,\,
\frac{\xi}{|\zeta|}
K_1^{-1}\qty(
\frac{\xi}{\sqrt{2}\,\zeta^2}
)
-\frac{1}{2}
}
.
\end{align}
As with the elastic cross section, we split the domain of integration,
\begin{align}
\frac{\sigma_{\T}}{\sigma_{\ell = 0}^\U}
&\approx
\int_{0}^{\ell_{\T}^\WKB} 
\dd \ell\ 
\qty(\ell + \frac{1}{2})
\sin^2\dot{\theta}_{\ell}(k)
+
\int_{\ell_{\T}^\WKB}^\infty 
\dd \ell\ 
\qty(\ell + \frac{1}{2})\ 
\sin^2\dot{\theta}_{\ell}(k)\nonumber\\[1ex]
&\approx
\frac{1}{4}\, \ell_{\T}^\WKB(\ell_{\T}^\WKB + 1)
+
\int_{\ell_{\T}^\WKB}^\infty 
\dd \ell\ 
\qty(\ell + \frac{1}{2})\ 
\qty[\dot{\theta}_{\ell}(k)]^2
,
\end{align}
where, as before, we replaced $\sin^2\dot{\theta}_{\ell}\to 1/2$. Using the $\ell$-derivative of \cref{eq:PS_Y_WKB_Approx}, we obtain
\begin{align}
\frac{\sigma_{\T}}{\sigma_{\ell = 0}^\U}
\approx
\frac{1}{4}\, \ell_{\T}^\WKB(\ell_{\T}^\WKB + 1)
+
\frac{\zeta^2}{2}\,
K_{\T}\qty[
\qty(\ell_{\T}^{\WKB} + \frac{1}{2})
\frac{|\zeta|
}{\xi}
]
,
\end{align}
where
\begin{align}\label{eq:K_T_def}
K_{\T}(z) = z^2 \qty{K_0(z)K_2(z) - \qty[K_1(z)]^2}
.
\end{align}
\paragraph{Viscosity cross section:} We define
\begin{align}
\qty[2\dot{\theta}_{\ell_{\V}^\WKB + \frac{1}{2}}^{\rm WP}(k)]^2 \equiv \frac{1}{2},
\end{align}
where the factor of 2 and the shift by 1/2 arise from~\cref{eq:V_CS_WKB}. Solving for $\ell_{\V}^{\WKB}(\zeta,\xi)$ gives
\begin{align}
\ell_{\V}^{\WKB}(\zeta,\xi)
=
\max\qty{
0,
\frac{\xi}{|\zeta|}
K_1^{-1}\qty(
\frac{\xi}{2\sqrt{2}\,\zeta^2}
)
- 1 
}
.
\end{align}
Once again, we split the domain of integration,
\begin{align}
\frac{\sigma_{\V}}{\sigma_{\ell = 0}^\U}
&\approx
\frac{1}{2}
\int_{0}^{\ell_{\V}^\WKB} 
\dd \ell\ 
(\ell + 1) 
\sin^2[2\dot{\theta}_{\ell + \frac{1}{2}}(k)]
+
\frac{1}{2}
\int_{\ell_{\V}^\WKB}^\infty 
\dd \ell\ 
(\ell + 1) 
\sin^2[2\dot{\theta}_{\ell + \frac{1}{2}}(k)]\nonumber\\[1ex]
&\approx
\frac{1}{8}\ell_{\V}^\WKB(\ell_{\V}^\WKB + 2)
+
\frac{1}{2}
\int_{\ell_{\V}^\WKB}^\infty 
\dd \ell\ 
(\ell + 1)
\qty[2\dot{\theta}_{\ell+\frac{1}{2}}(k)]^2
,
\end{align}
where, as before, we replaced $\sin^2[2\dot{\theta}_{\ell+1/2}]\to 1/2$. Using the $\ell$-derivative of \cref{eq:PS_Y_WKB_Approx} we obtain,
\begin{align}
\frac{\sigma_{\V}}{\sigma_{\ell = 0}^\U}
\approx
\frac{1}{8}\ell_{\V}^\WKB(\ell_{\V}^\WKB + 2)
+
\zeta^2\,
K_{\T}\qty[
(\ell_{\V}^{\WKB} + 1)
\frac{
|\zeta|
}{\xi}
]
.
\end{align}

\clearpage

\end{appendices}

\clearpage
\phantomsection
\addcontentsline{toc}{section}{References} 
\bibliography{biblio.bib}
\bibliographystyle{JHEP.bst}

\end{document}